\documentclass[%
reprint,
11 pt,
amsmath,amssymb,
aps,
pra,
nofootinbib,
onecolumn,
preprintnumbers,
superscriptaddress,
tightenlines
]{revtex4-1}
\usepackage[paperwidth=210mm,paperheight=297mm,centering,hmargin=2.3cm,vmargin=2.3cm]{geometry}
\usepackage{dsfont}
\usepackage{braket}
\usepackage{bm}
\usepackage{dcolumn}
\usepackage{subcaption}
\usepackage{graphicx}
\usepackage[colorlinks=true,citecolor=magenta,linkcolor=blue
]{hyperref} 
\usepackage{multirow}
\newcommand{\obar}{\underline{1}}
\newcommand{\nui}{\nu_{\underline{1}}}
\newcommand{\num}{\nu_0}
\newcommand{\nuo}{\nu_1}
\newcommand{\hnui}{\hat{\nu}_{\underline{1}}}
\newcommand{\hnum}{\hat{\nu}_0}
\newcommand{\hnuo}{\hat{\nu}_1}
\newcommand{\ladder}{\Gamma}
\newcommand{\ApBp}{ A(\obar)^\dagger \cdot B(1)^\dagger}
\newcommand{\BpAp}{ B(\obar)^\dagger \cdot A(1)^\dagger}

\newcommand{\kket}[1]{\left| \left| #1 \right\rangle \right.}

\newcolumntype{C}[1]{>{\centering\let\newline\\\arraybackslash\hspace{2pt}}m{#1}}
\newcolumntype{R}[1]{>{\centering\let\newline\\\arraybackslash\hspace{0pt}}m{#1}}
\newcommand{\goldieotimes}{%
  \mathop{\mathchoice{\textstyle\bigotimes}{\bigotimes}{\bigotimes}{\bigotimes}}%
}

\usepackage{soul}

\begin{document}

\preprint{UMD-PP-022-11}

\title{Loop-string-hadron formulation of an SU(3) gauge theory\\with dynamical quarks}

\author{Saurabh V. Kadam}
 \email{ksaurabh@umd.edu}
\affiliation{%
 Maryland Center for Fundamental Physics, University of Maryland, College Park, MD 20742, USA
}%
\author{Indrakshi Raychowdhury}
 \email{indrakshir@goa.bits-pilani.ac.in}
\affiliation{Department of Physics, BITS-Pilani,
K K Birla Goa Campus, Zuarinagar, Goa 403726, India}%
\author{Jesse R. Stryker}
 \email{strykerj@umd.edu}
\affiliation{%
 Maryland Center for Fundamental Physics, University of Maryland, College Park, MD 20742, USA
}%

\date{\today}

\begin{abstract}
    Towards the goal of quantum computing for lattice quantum chromodynamics, we present a loop-string-hadron (LSH) framework in 1+1 dimensions for describing the dynamics of SU(3) gauge fields coupled to staggered fermions.
    This novel framework was previously developed for an SU(2) lattice gauge theory in $d\leq3$ spatial dimensions and its advantages for classical and quantum algorithms have thus far been demonstrated in $d=1$. 
    The LSH approach uses gauge invariant degrees of freedoms such as loop segments, string ends, and on-site hadrons, it is free of all nonabelian gauge redundancy, and it is described by a Hamiltonian containing only local interactions.
    In this work, the SU(3) LSH framework is systematically derived from the reformulation of Hamiltonian lattice gauge theory in terms of irreducible Schwinger bosons
    , including the addition of staggered quarks.
    Furthermore, the superselection rules governing the LSH dynamics are identified directly from the form of the Hamiltonian.
    The SU(3) LSH Hamiltonian with open boundary conditions has been numerically confirmed to agree with the completely gauge-fixed Hamiltonian, which contains long-range interactions and does not generalize to either periodic boundary conditions or to $d>1$.
\end{abstract}

\maketitle
\tableofcontents

\section{Introduction
\label{sec: intro}
}

\noindent
Quantum technology is expected to revolutionize various problems---financial forecasting, materials research, and precision measurements, to name a few---but there are only a handful of applications for which rigorous arguments have actually been made implying a quantum speedup.
In the natural sciences, the lone such application is the simulation of quantum mechanical systems~\cite{Feynman:1981tf,doi:10.1126/science.273.5278.1073}.
The simulation of quantum-mechanical systems is not entirely new, however;
enormous amounts of power are already dedicated to classical supercomputers running simulations of chemical and physical systems in which quantum effects are important.
The problem is that simulations on classical machines often face limitations in simulating certain systems~\cite{Troyer:2004ge} or with scaling up the problem size.
In high-energy physics, it is lattice quantum field theory---in particular, lattice quantum chromodynamics (QCD)---where simulations are most prominent and where there is opportunity for a quantum boost \cite{Bauer:2022hpo}.
Classical simulation protocols, however, do not immediately carry over onto quantum hardware;
the so-called second quantum revolution has thus started a race to understand how to design quantum simulation protocols in theory and to subsequently bring theoretical proposals into alignment with laboratory reality.

Full-blown lattice QCD is marvelously rich in technical features, including a multitude of quark flavors and masses, three spatial dimensions, confinement of quarks, and spontaneous breaking of chiral symmetry, but perhaps the single most defining feature of QCD is its nonabelian, SU(3) gauge symmetry.
Understanding how to implement SU(3)-symmetric matter and interactions, therefore, will be one of the essential milestones on the path to quantumly simulating QCD. 
The aim of the present work is to advance our understanding of how SU(3) gauge theories can be formulated for quantum computers.

Some of the most promising opportunities for a quantum boost in lattice field theory include calculations whose path integrals suffer from so-called ``sign problems.''
While sign problem solutions may be found for special cases~\cite{Grabowska:2012ik}, or in particular parameter regimes~\cite{deForcrand:2002hgr}, a general-purpose solution to a wide variety of problems of interest has not been found and may be nonexistent~\cite{Troyer:2004ge}.
Scenarios in which sign problems arise include nonzero chemical potential, topological terms in the Lagrangian, and far-from-equilibrium and real-time dynamics.
Traditional lattice (gauge) field theory \cite{Wilson:1974sk} has been based on importance sampling of path integrals in imaginary time.
But the Monte Carlo importance sampling approach breaks down when the Euclidean action acquires imaginary contributions or when considering finite intervals of real time.
In contrast, the Hamiltonian formalism, expressed in terms of vector spaces, linear operators, time evolution in a Schr\"{o}dinger (or Heisenberg or Dirac) picture, operations on states, and measurement probabilities does not obviously suffer from a sign problem.
Instead, it suffers from the exponential growth of Hilbert space -- regardless of what terms may be in the Hamiltonian.
Quantum simulators, themselves featuring exponentially large Hilbert spaces and being naturally expressed in the Hamiltonian framework, may be the key to opening up this method of calculation. There have been continuous efforts over the past decade
\cite{Banerjee:2012pg, Banerjee:2012xg, Zohar:2012xf, Tagliacozzo:2012df, Zohar:2015hwa, Tagliacozzo:2012vg, Zohar:2011cw,Aidelsburger:2021mia, Mazza:2011kf, Gonzalez-Cuadra:2017lvz, Zohar:2016iic, Kasper:2016mzj, Muschik:2016tws,Atas:2022dqm, Farrell:2022wyt, Farrell:2022vyh, Armon:2021uqr, Carena:2022kpg, Martinez:2016yna, Davoudi:2019bhy, Klco:2018kyo, Klco:2019evd,Davoudi:2021ney,Raychowdhury:2018osk, Dasgupta:2020itb, Mil:2019pbt, Semeghini:2021wls, Yang:2020yer, Atas:2021ext, Ciavarella:2021nmj, Paulson:2020zjd,Riechert:2021ink, Halimeh:2019svu, Halimeh:2021vzf, Zhou:2021kdl,Stryker:2021asy,ARahman:2022jks,Yamamoto:2022jnn,Kane:2022ejm}
toward this research goal. 

Quantum simulation of field theory Hamiltonians raises important and interesting questions surrounding truncation \cite{Kessler:2003cv,Tong:2021rfv} and field discretization \cite{Klco:2018zqz}, but for gauge theories in particular there has been a flurry of investigations into different \emph{formulations} of a given model \cite{Davoudi:2020yln}.%
\footnote{
We use ``formulation'' loosely to mean a defined set of local degrees of freedom and their associated values, any relevant operator identities or constraints on the Hilbert space, and a set of operators or rules for constructing interactions between the degrees of freedom.
A given formulation may often be associated with one or more particular choices of basis.}
These formulations include the original Kogut-Susskind formulation~\cite{Kogut:1974ag}, dual or magnetic variables \cite{Horn:1979fy,Ukawa:1979yv,Kaplan:2018vnj,Haase:2020kaj,Bauer:2021gek}, purely fermionic formulations (in 1D space) \cite{Hamer:1981yq,Banuls:2017ena,Farrell:2022wyt,Atas:2022dqm}, local-multiplet bases \cite{Banuls:2017ena,Klco:2019evd,Ciavarella:2021nmj}, quantum link models \cite{Chandrasekharan:1996ih}, tensor formulations \cite{Bazavov:2015kka,Meurice:2021bcz,Meurice:2020pxc}, prepotential (Schwinger boson) formulations \cite{Mathur:2004kr,Mathur:2007nu,Anishetty:2009ai,Anishetty:2009nh,Mathur:2010wc,Raychowdhury:2013rwa,Raychowdhury:2018tfj}, the loop-string-hadron (LSH) formulation \cite{Raychowdhury:2019iki,Raychowdhury:2018osk} derived from prepotentials, and more \cite{Fontana:2020xzp}.
The availability of different formulations is a boon to the lattice community; some formulations can be more efficient for different parameter regimes \cite{Haase:2020kaj,Bauer:2021gek}, and alternate formulations may prove useful for consistency checks in the post-classical computing regime.
For these reasons, it is important to understand the advantages and drawbacks to a given formulation.

Much can be learned about the challenges that will be encountered in QCD by first considering gauge groups simpler than SU(3), namely U(1) and SU(2).
Understanding SU(2) is especially crucial as it features the nonabelian interactions that are so characteristic of QCD, and the representation theory is nontrivial.
In the case of SU(2), the recently proposed loop-string-hadron formulation was suggested to have several attractive features from the perspective of quantum simulation.
The LSH formulation is derived from the Schwinger boson or prepotential formulation, and in certain respects represents the culmination of the work that has gone into the latter.
The LSH Hilbert space and Hamiltonian have been developed at length for the SU(2) gauge group, as well as matter in the form of one flavor of staggered fermions.
Over time, benefits of the LSH framework have been substantiated~\cite{Davoudi:2020yln} and discovered~\cite{Davoudi:2022xmb}, although there is much work left to do -- especially in multidimensional space, where it is already known that the qubit savings could break down \cite{Raychowdhury:2018osk}.
Still, these preliminary investigations encourage the continued exploration into the LSH framework.

One of the exciting possibilities is that the LSH approach will generalize to SU(3) and continue to offer computational advantages -- perhaps even yielding greater savings.
The original LSH-formulated theory was derived from the earlier Schwinger boson or ``prepotential'' formulations of SU(2) lattice gauge theories.
Schwinger bosons have already been generalized to SU(3) [and even SU(N)] lattice gauge theories.
The transition from SU(2) Schwinger bosons to SU(3) comes with considerable new technical challenges, but we have discovered that an SU(3) LSH framework can be constructed using essentially the same building blocks found in SU(2) -- at least for the model explicitly considered, i.e., Yang-Mills in 1+1 dimensions coupled to one flavor of staggered quarks.
A recent algorithmic study of the SU(2) analogue of this model confirms cost saving advantages with the LSH formulation \cite{Davoudi:2022xmb}, which we can speculate to be even larger for SU(3) based purely on the structure of the Hamiltonian.

During the course of this work, the development of simulation protocols for SU(3) gauge theories in other formulations has picked up steam, but the topic remains very much in its infancy.
The first study to ever explore algorithms for U(1), SU(2), and SU(3) lattice gauge theories was Ref.~\cite{Byrnes:2005qx}, while a more contemporary and exhaustive approach has been worked out more recently in Ref.~\cite{Kan:2021xfc}.
The latter has placed an upper bound of $\sim 10^{49}$ on the gate count cost of computing transport coefficients in QCD, indicating a vast gap between revolutionary quantum algorithms for QCD and hardware reality.
On the other hand, recent studies with more focus on near-term applications have ported small SU(3) systems onto existing hardware \cite{Ciavarella:2021nmj,Atas:2022dqm,Farrell:2022vyh,Farrell:2022wyt}, although most of these benefit from the ability to completely integrate out gauge fields in 1D space.
In this article, we develop the LSH formulation of SU(3) lattice gauge theory with staggered quarks in 1D space, retaining the gauge fields that are needed in higher dimensions while also postponing the complications of chromomagnetic interactions to future work.\footnote{
Following the original LSH construction~\cite{Raychowdhury:2018osk,Raychowdhury:2019iki}, a similar framework was proposed and extended to SU(3)~\cite{P:2019qdq}.
This variation uses more matter quantum numbers (tied by additional constraints), and has a different point-splitting scheme for accommodating matter.
Furthermore, that framework's Hamiltonian has not yet been made available and demonstrated to be equivalent numerically, nor has it been applied to quantum simulation with demonstrable cost-saving advantages.
For this reason, we see it necessary to revisit the LSH approach to lattice gauge theory Hilbert spaces and generalize LSH as it was originally conceived in Ref.~\cite{Raychowdhury:2019iki} to SU(3).
}
We will begin in Sec.~\ref{sec: results} by directly going into a concise summary of the LSH theory's definitions and Hamiltonian, plus a numerical confirmation of its ability to produce the spectrum expected from a more conventional, ``purely fermionic'' formulation.
The actual derivation of the LSH Hilbert space and Hamiltonian from SU(3) Schwinger bosons is deferred to Sec.~\ref{sec: LSH-framework}, for it turns out the calculations involved can be long and technical yet they have little bearing on applying the resultant LSH framework.
Before that derivation, Sec. \ref{sec: methods1} first presents the conventional Kogut-Susskind formulation and the equivalent prepotential formulation from which the LSH formulation is derived.
Our conclusions in Sec.~\ref{sec: discussion} summarize the SU(3) LSH formulation and provide perspectives on how this work contributes to the program of developing quantum-technological applications for lattice QCD.


\section{Results
\label{sec: results}
}
\noindent
This section presents the key ingredients of the LSH formalism for SU(3) gauge theory defined on a one-dimensional spatial lattice, namely, the Hilbert space, diagonal and off-diagonal operators acting within that Hilbert space, and the Hamiltonian constructed out of these operators. The purpose of summarizing the results here is to encourage non-technical readers to appreciate the salient features of this formalism that make it helpful for developing algorithms based on either classical or quantum technologies.
The results close with a numerical benchmark confirming that the formulation introduced herein does in fact reproduce the expected energy eigenvalues.
\subsection{LSH occupation-number basis
\label{subsec: LSH basis}
}

\begin{figure}[t]
    \includegraphics[scale=0.99]{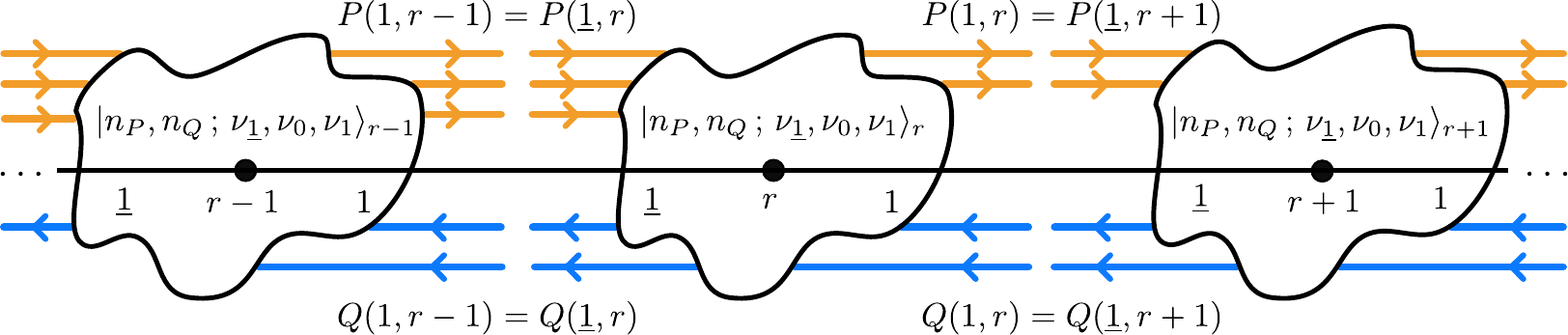}
    \caption{
    Schematic representation of LSH degrees of freedom in an occupation number basis, with on-site states that satisfy the Abelian Gauss's laws.
    Here, filled circles labeled by $r-1$, $r$, and $r+1$ depict the lattice sites at those positions. The left and right directions from each lattice site are denoted by $\obar$ and $1$, respectively. The orange and blue lines denote the $P$-type and $Q$-type gauge flux units, respectively, and they are related to the on-site LSH quantum numbers at each site through Eqs.~\eqref{eq: P1Q1} and~\eqref{eq: P2Q2}. The arrows on them indicate the direction in which the corresponding gauge flux passes through a site, rightward $(\obar \rightarrow 1)$ of $P$-type and leftward $(1 \rightarrow \obar)$ for $Q$-type gauge flux. The LSH basis states satisfy the Abelian Gauss's law constraints defined in Eq.~\eqref{eq: AGL} which requires that the number of flux units of each $P$- and $Q$-type of gauge flux to be conserved over each link connecting two sites, as shown in the figure.\label{fig: AGL_pict}}
\end{figure}

We consider an SU(3) lattice gauge theory coupled to staggered fermionic matter on a one-dimensional spatial lattice with $N$ lattice points, where $N$ is even. 
The LSH Hilbert space for this theory is distinct from other formulations in that it is strictly SU(3)-invariant.
It is spanned by an occupation-number (also called ``electric,'' ``strong-coupling,'' or ``flux'') basis that is  characterized by two bosonic  or pure gluonic excitations $(n_P ,n_Q)$ and three fermionic or matter-gauge degrees of freedom $(\nui , \num , \nuo )$ at each site of the (staggered) lattice, $r$, as depicted in the blobs of Fig.~\ref{fig: AGL_pict}.
The bosonic quantum numbers $(n_P ,n_Q)$ represent the oriented flow of gluonic flux, i.e., pure gauge flux units pointing in either the rightward $(\obar \rightarrow 1)$ or leftward $(1 \rightarrow \obar)$ direction through the site. The fermionic quantum numbers $(\nui,\num,{\nuo})$ are associated with the presence of quarks coupled to gauge flux which may flow into or out of the site, and to the $1$ side or the $\obar$ side or both.
The on-site, orthonormal LSH states in the electric basis are thus characterized as
\begin{equation}
    \ket{n_P, n_Q\,;\,\nu_{\obar},{\nu_0},\nu_1}_r , \qquad n_P, n_Q \in \{0,1,2,\cdots \} , \qquad \nu_{\obar}, \nu_0 , \nu_1 \in \{ 0,1 \},
    \label{eq: LSH-state}
\end{equation}
at any site $r=1,2,\cdots,N$.
Being a rank two group, an SU(3) irreducible representation (irrep) is characterized by two ``indices'' (non-negative integers) $P$ and $Q$.
The electric flux configurations to either side of a site are identified by these irreps, hence there is $(P(\obar,r),Q(\obar,r))$ to the $\obar$ side of site $r$ and $(P(1,r),Q(1,r))$ to the $1$ side.
The irrep quantum numbers to either side $\obar$ or $1$ of site $r$ are related to the on-site LSH occupation numbers via
\begin{align}
        P(\obar,r) = n_P(r) +  \nu_0(r)\left(1-\nu_1 (r) \right),~&~Q(\obar,r)= n_Q(r) +  \nu_{\obar}(r)\left(1-\nu_0 (r) \right), \label{eq: P1Q1}
       \\
        P(1,r)= n_P(r) +  \nu_1(r)\left(1-\nu_0 (r) \right), ~&~ Q(1,r) = n_Q(r) +  \nu_0(r)\left(1-\nu_{\obar} (r) \right),\label{eq: P2Q2}
\end{align}
where $n_P(r)$, $n_Q(r)$, $\nu_{\obar}(r)$, $\nu_0(r)$, and $\nu_1(r)$ denote the values of their corresponding on-site LSH quantum numbers at site $r$.  
As indicated in Eqs.~\eqref{eq: P1Q1}-\eqref{eq: P2Q2}, in addition to the bosonic excitations $n_P$ and $n_Q$, the matter content characterized by $\nu_{\obar}$, $\nu_0$, and $\nu_1$ acts as a source or sink for one or both directions of gauge flux; this is quantified by the contributions of the form $\nu_f (1-\nu_{f'})$.
The LSH flux basis for the lattice as a whole is essentially a tensor product basis, except for the presence of certain Abelian Gauss's law constraints, that is a defining feature of the LSH framework. The Abelian Gauss's laws relate the SU(3) irreps at both ends of a link connecting sites $r$ and $r+1$ as
\begin{align}
    P(1,r) = P(\obar,r+1)  \quad\text{and}\quad
    Q(1,r) = Q(\obar,r+1).
    \label{eq: AGL}
\end{align}
%
The set of constraints given in Eq.~\eqref{eq: AGL} guarantees the conservation of flux units of both directions (combining contributions from both loop segments and string ends) in both $(1 \rightarrow \obar)$ and $(\obar \rightarrow 1)$ directions along the link connecting sites $r$ and $r+1$, see Fig.~\ref{fig: AGL_pict} for a pictorial representation.
The tensor product space of local Hilbert spaces from every site $r$, together with the Abelian Gauss's law constraints, is equivalent to the gauge-invariant subspace in the Kogut-Susskind formulation and describes extended strings and hadron excitations distributed throughout the lattice.
Closed flux loops are possible on the one-dimensional lattice if the boundary conditions are periodic.

\subsection{\label{subsec: LSH_operators}LSH number operators, diagonal functions, and ladder operators}

In the same way that the Hilbert space of the lattice is constructed from on-site local Hilbert spaces, the Hamiltonian for the lattice is constructed in terms of on-site operators acting on the local Hilbert spaces.
These operators are the LSH number and ladder operators, to be defined in this section.

The LSH number operators are diagonal in the LSH flux basis defined in the previous section, with eigenvalues corresponding to the bosonic and fermionic occupation numbers. Explicitly, the bosonic and fermionic number operators at site $r$ are given by
\begin{align}
    \hat{n}_P (r)   &= \sum_{n_P, n_Q,\nu_{\obar} , \nu_0 , \nu_1} n_P (r) \ket{n_P,n_Q\,;\,\nu_{\obar},\nu_0,\nu_1}\bra{n_P,n_Q\,;\,\nu_{\obar},\nu_0,\nu_1}_r,
    \label{eq: LSH-diag-nP}   \\
    \hat{n}_Q  (r) &= \sum_{n_P, n_Q,\nu_{\obar} , \nu_0 , \nu_1} n_Q(r) \ket{n_P,n_Q\,;\,\nu_{\obar},\nu_0,\nu_1} \bra{n_P,n_Q\,;\,\nu_{\obar},\nu_0,\nu_1}_r,
    \label{eq: LSH-diag-nQ}\\
    \hnui (r) &= \sum_{n_P, n_Q,\nu_{\obar} , \nu_0 , \nu_1} \nu_{\obar}(r) \ket{n_P,n_Q\,;\,\nu_{\obar},\nu_0,\nu_1} \bra{n_P,n_Q\,;\,\nu_{\obar},\nu_0,\nu_1}_r,
    \label{eq: LSH-diag-nu1bar}\\
    \hnum (r) &= \sum_{n_P, n_Q,\nu_{\obar} , \nu_0 , \nu_1} \nu_0(r) \ket{n_P,n_Q\,;\,\nu_{\obar},\nu_0,\nu_1} \bra{n_P,n_Q\,;\,\nu_{\obar},\nu_0,\nu_1}_r,
    \label{eq: LSH-diag-nuo}\\
    \hat{\nu}_{1} (r) &= \sum_{n_P, n_Q,\nu_{\obar} , \nu_0 , \nu_1} \nu_1(r) \ket{n_P,n_Q\,;\,\nu_{\obar},\nu_0,\nu_1} \bra{n_P,n_Q\,;\,\nu_{\obar},\nu_0,\nu_1}_r,
    \label{eq: LSH-diag-nu1}
\end{align}
where the `$(r)$' arguments to the indices under each summation symbol have been left implicit.
In addition to the LSH number operators, there are other frequently-occurring operators that are diagonal in the LSH flux basis worth mentioning.
The first are the irrep quantum numbers, i.e. the Hilbert space operators corresponding to Eqs.~\eqref{eq: P1Q1}-\eqref{eq: P2Q2}:
\begin{align}
        \hat{P}(\obar,r) = \hat{n}_P(r) +  \hat{\nu}_0(r)\left(1-\hat{\nu}_1 (r) \right),~&~\hat{Q}(\obar,r)= \hat{n}_Q(r) +  \hat{\nu}_{\obar}(r)\left(1-\hat{\nu}_0 (r) \right), \label{eq: P1Q1_operators}
       \\
        \hat{P}(1,r)= \hat{n}_P(r) +  \hat{\nu}_1(r)\left(1-\hat{\nu}_0 (r) \right), ~&~ \hat{Q}(1,r) = \hat{n}_Q(r) +  \hat{\nu}_0(r)\left(1-\hat{\nu}_{\obar} (r) \right).
        \label{eq: P2Q2_operators}
\end{align}
Secondly, we will use the term ``diagonal functions'' to refer to
any operator that can be written as a closed-form function of the number operators in Eqs.~\eqref{eq: LSH-diag-nP}-\eqref{eq: LSH-diag-nu1}.
For example, if $F(m,n)\equiv \sqrt{m+n+1}$ for non-negative integers $m$ and $n$, there exists a corresponding diagonal function $F(\hat{n}_P(r),\hat{n}_Q(r))$ given by
\begin{equation*}
    \sqrt{\hat{n}_P(r)+\hat{n}_Q(r)+1} \equiv \sum_{n_P, n_Q, \nu_{\obar} , \nu_0 , \nu_1} \ket{n_P,n_Q\,;\,\nu_{\obar},\nu_0,\nu_1} \bra{n_P,n_Q\,;\,\nu_{\obar},\nu_0,\nu_1}_r \sqrt{n_P+n_Q+1} .
\end{equation*}

The lowering (raising) ladder operator for each LSH quantum number is defined for each site $r$, and it lowers (raises) the value of the corresponding on-site LSH quantum number by 1 in a way that is  consistent with the bosonic or fermionic statistics of the corresponding quantum number. The bosonic lowering operators are defined by
\begin{align}
    \hat{\ladder}^{}_{P} (r) &= \sum_{n_P=1}^\infty \ \sum_{n_Q,\nu_{\obar} , \nu_0 , \nu_1} \ket{n_P-1, n_Q\,;\,\nu_{\obar},\nu_0,\nu_1} \bra{n_P,n_Q\,;\,\nu_{\obar},\nu_0,\nu_1}_r, \label{eq: Lp-local} \\
    \hat{\ladder}^{}_{Q} (r) &= \sum_{n_Q=1}^\infty \ \sum_{n_P,\nu_{\obar} , \nu_0 , \nu_1} \ket{n_P,n_Q-1\,;\,\nu_{\obar},\nu_0,\nu_1} \bra{n_P,n_Q\,;\,\nu_{\obar},\nu_0,\nu_1}_r, \label{eq: Lq-local}
\end{align}
whereas, the fermionic lowering operators at site $r$ are given by
\begin{align}
    \hat{\chi}_{\obar}(r) &=  \left( \prod_{r'=1}^{r-1} (-1)^{\hat{\nu}_{\obar}(r')+\hat{\nu}_0(r')+\hat{\nu}_1(r')} \right)\sum_{n_P , n_Q , \nu_0 , \nu_1 } \ket{n_P,n_Q\,;\,0 ,\nu_0 ,\nu_1} \bra{n_P,n_Q\,;\,1 ,\nu_0 ,\nu_1}_r, \label{eq: chi-1bar-local} \\
    \hat{\chi}_0 (r) &=  \left( \prod_{r'=1}^{r-1} (-1)^{\hat{\nu}_{\obar}(r')+\hat{\nu}_0(r')+\hat{\nu}_1(r')} \right)\sum_{n_P , n_Q , \nu_{\obar} , \nu_1 } \ket{n_P,n_Q\,;\,\nu_{\obar} ,0 ,\nu_1} \bra{n_P,n_Q\,;\,\nu_{\obar} ,1 ,\nu_1}_r (-1)^{\nu_{\obar}}, \label{eq: chi-o-local} \\
    \hat{\chi}_{1}(r) &=  \left( \prod_{r'=1}^{r-1} (-1)^{\hat{\nu}_{\obar}(r')+\hat{\nu}_0(r')+\hat{\nu}_1(r')} \right)\sum_{n_P , n_Q , \nu_{\obar} , \nu_0 } \ket{n_P,n_Q\,;\,\nu_{\obar} ,\nu_0 ,0} \bra{n_P,n_Q\,;\,\nu_{\obar} ,\nu_0 ,1}_r (-1)^{\nu_{\obar} + \nu_0}, \label{eq: chi_1_local} 
\end{align}
with the corresponding raising operators $\hat{\ladder}_P^\dagger(r)$, $\hat{\ladder}_Q^\dagger(r)$,  $\hat{\chi}_{\obar}^\dagger(r)$,  $\hat{\chi}_0^\dagger(r)$, and  $\hat{\chi}_{1}^\dagger(r)$, being their conjugates.
Here, the factors containing $(-1)^{\hat{\nu}_f(r')}= 1-2\hat{\nu}_f(r')$ depend on the fermionic quantum numbers at lattice sites from $r'=1$ to $r-1$ such that $\hat{\chi}_f(r)$ operators satisfy the canonical anticommutation relations:
\begin{align}
    \label{eq: LSH-anticomm-vanishing}
    \left\{ \hat{\chi}_{f} (r), \hat{\chi}_{f'} (r') \right\} =\bigl\{ \hat{\chi}_{f}^{\dagger} (r), \hat{\chi}_{f'}^{\dagger} (r') \bigr\} &= 0 , \\
    \label{eq: LSH-anticomm-kronecker}
   \bigl\{ \hat{\chi}_{f} (r), \hat{\chi}_{f'}^{\dagger} (r') \bigr\} &= \delta_{f \, f'} \delta_{r\,r'},
\end{align}
for all possible choices of $r$, $r'$, $f$, and $f'$.
Note that the sign factors in $\hat{\chi}_f(r)$, despite being nonlocal in their definitions, are of little consequence on the one-dimensional lattice since the contributions coming from all but two sites will cancel upon substitution into the Hamiltonian, as will be seen in Sec.~\ref{subsec: LSH-ham}.

The $\hat{\ladder}_l(r)$ and $\hat{\chi}_f(r)$ operators and their adjoints are ``normalized'' in the sense that $\hat{\mathcal{O}}^\dagger \hat{\mathcal{O}}$ is a (normalized) projection operator when $\hat{\mathcal{O}}$ is any of these ten operators.
The raising operators can be used to express the LSH flux basis states simply as follows:
\begin{equation}
    \ket{n_P,n_Q\,;\,\nu_{\obar},\nu_0,\nu_1} =
    ( \hat{\ladder}_{P}^{\dagger} ) ^{n_P}
    ( \hat{\ladder}_{Q}^{\dagger} ) ^{n_Q}
    ( \hat{\chi   }_{\obar}^{\dagger} ) ^{\nu_{\obar}}
    ( \hat{\chi   }_{0}^{\dagger} ) ^{\nu_0}
    ( \hat{\chi   }_{1}^{\dagger} ) ^{\nu_1} \ket{0,0\,;\,0,0,0} ,
    \label{eq: LSH-basis-ket}
\end{equation}
where $\ket{0\\,0\,;\,0,0,0}$ denotes a state with zero occupation of all of the loop-string-hadron excitations. Note that, for notational brevity, 
when only one site is under consideration, as in Eq.~\eqref{eq: LSH-basis-ket}, we may suppress the site index $r$ from LSH states, LSH quantum numbers, and ladder operators. Lastly, a recurring structure that helps to concisely express local SU(3)-invariant operators is a ``conditional'' (bosonic) ladder operator.
Conditional ladder operators have the form
\begin{equation}
    ( \hat{\ladder}_l )^{  \hat{\nu}_f} \equiv \hat{\nu}_f \hat{\ladder}_l + ( 1 - \hat{\nu}_f ), \qquad ( \hat{\ladder}_l^{\dagger} )^{  \hat{\nu}_f} \equiv \hat{\nu}_f \hat{\ladder}_l^{\dagger} + ( 1 - \hat{\nu}_f ), \qquad l \in \{P, Q\}, \qquad f\in\{\obar,0,1\}.
    \label{eq: cond-ladd-operator}
\end{equation}
In words, this says ``apply $\hat{\ladder}_l^{(\dagger)}$, but only if $\nu_f = 1$.''

\subsection{The LSH Hamiltonian
\label{subsec: LSH-ham}
}
The lattice Hamiltonian for interacting quarks on a one dimensional lattice consists of three distinct parts:
\begin{equation}
    H = H_M + H_E + H_{I} ,
    \label{eq: ham-terms}
\end{equation}
where $H_M$ is the matter self-energy, $H_E$ is the chromoelectric energy, and $H_{I}$ is the gauge-matter interaction.\footnote{The Hamiltonian in Eq.~(\ref{eq: ham-terms}) is written in a dimensionless form. To obtain the traditional, dimensionful lattice Hamiltonian, one should multiply Eq.~(\ref{eq: ham-terms}) by $\frac{g_0^2 a_s}{2}$, replace $\mu\to\frac{2m_0}{g_0^2 a_s} $, and replace $x\to\frac{1}{g_0^2 a_s^2}$; the parameters $a_s$, $g_0$, and $m_0$ represent the spatial lattice spacing, bare gauge coupling, and bare fermion mass, respectively \cite{Hamer:1981yq}.}
The Hamiltonian is expressed explicitly in terms of the on-site LSH operators. The mass and electric Hamiltonians, $H_M$ and $H_E$, are diagonal operators in the LSH flux basis, depending only on the on-site LSH number operators defined in Eqs.~\eqref{eq: LSH-diag-nP}-\eqref{eq: LSH-diag-nu1}.
The matter self-energy depends on the total number of quarks at each site as expressed by
\begin{equation}
    H_M = \sum_{r=1}^{N} H_M(r) = \sum_{r=1}^{N} \mu (-1)^r \big( \hnui (r) +  \hnum (r) + \hat{\nu}_{1} (r) \big) ,
    \label{eq: HM}
\end{equation}
for both open and periodic boundary conditions, and $\mu$ is the fermionic mass parameter.
$H_E$ measures the electric energy along the links as quantified by the quadratic Casimir associated with the irreps. In terms of the SU(3) irrep quantum numbers on a link connecting sites $r$ and $r+1$, the electric Hamiltonian is expressed by
\begin{equation}
    H_E = \sum_{r=1}^{N'} H_E(r) = \sum_{r} \frac{1}{3} \Big( \hat{P}(1,r)^2 + \hat{Q}(1,r)^2 + \hat{P}(1,r) \hat{Q}(1,r) \Big) + \hat{P}(1,r) + \hat{Q}(1,r) ,
    \label{eq: HE}
\end{equation}
where the number of links $N'$ equals $N-1$ ($N$) for open (periodic) boundary conditions.
The gauge-matter interaction or ``hopping Hamiltonian'' $H_{I}$, responsible for dynamics, is the most structurally rich part of the theory. It is expressed using the raising and lowering operators of Eqs.~(\ref{eq: Lp-local})-(\ref{eq: chi_1_local}) (in addition to the LSH number operators), such that it is strictly off-diagonal in the flux basis.
The total gauge-matter interaction energy can be written as a sum of its localized couplings that describe the fermionic ``hopping'' between two neighboring lattice sites. Explicitly, 
\begin{equation}
    H_{I} = \sum_{r=1}^{N'} H_I(r,r+1),
\end{equation}
with
\begin{align}
H_I(r,r+1) &\equiv x \left[ 
    \hat{\chi}_{1}^\dagger ( \hat{\ladder}_P^\dagger )^{\hnum}
    \sqrt{ 1  - \hnum/(\hat{n}_P + 2)} \sqrt{ 1 - \hnui/(\hat{n}_P+\hat{n}_Q+3)}
    \ \right]_{r} \nonumber \\
    & \hspace{1.2cm} \otimes
    \left[
    \sqrt{ 1  + \hnum/(\hat{n}_P + 1)} \sqrt{ 1 + \hnui/(\hat{n}_P+\hat{n}_Q+2)} \,
    \hat{\chi}_{1} ( \hat{\ladder}_P^\dagger )^{1-\hnum}
    \right]_{r+1} \nonumber \\
    &\quad + x \left[
    \hat{\chi}_{\obar}^\dagger ( \hat{\ladder}_Q )^{1-\hnum}
    \sqrt{ 1 + \hnum/(\hat{n}_Q + 1)}\sqrt{ 1 + \hat{\nu}_{1}/(\hat{n}_P+\hat{n}_Q+2)}
    \ \right]_{r} \nonumber \\
    &\hspace{1.2cm} \otimes \left[
    \sqrt{ 1 - \hnum/(\hat{n}_Q + 2)} \sqrt{ 1 - \hat{\nu}_{1}/(\hat{n}_P+\hat{n}_Q+3)} \, \hat{\chi}_{\obar} ( \hat{\ladder}_Q )^{\hnum}
    \right]_{r+1} \nonumber \\
    &\quad + x \left[
    \hat{\chi}_0^\dagger ( \hat{\ladder}_P )^{1-\hat{\nu}_{1}} ( \hat{\ladder}_Q^\dagger )^{\hnui}
    \sqrt{ 1 + \hat{\nu}_{1}/({\hat{n}_P+1}}) \sqrt{ 1 - {\hnui}/({\hat{n}_Q+2})}
    \ \right]_{r} \nonumber \\
    & \hspace{1.2cm} \otimes \left[
    \sqrt{ 1 - {\hat{\nu}_{1}}/({\hat{n}_P+2}}) \sqrt{ 1 + {\hnui}/({\hat{n}_Q+1}}) \,
    \hat{\chi}_0 ( \hat{\ladder}_P )^{\hat{\nu}_{1}} ( \hat{\ladder}_Q^\dagger )^{1-\hnui}
    \right]_{r+1} + \mathrm{H.c.},
     \label{eq: HI}
\end{align}
and $x$ being the interaction strength. Equation~\eqref{eq: HI} is a key result of this work and will be derived in Sec. \ref{sec: LSH-framework}. In Eq.~\eqref{eq: HI}, we have used the notation $\left[ \; \right]_{r}$ for brevity  to denote a product of operators all associated with site $r$.

\subsection{Global symmetry sectors of the SU(3) LSH formulation
\label{subsec: super-selection}
}
It is important to note the global symmetries of the LSH formulation that lead to a block-diagonalized structure of the LSH Hamiltonian. The gauge-invariant dynamics then remains confined into each block providing significant computational benefit. To identify these global symmetries, we notice that mass and electric Hamiltonians in Sec.~\ref{subsec: LSH-ham} are already diagonal in the LSH basis. On the other hand, the interaction Hamiltonian, although being non-diagonal, conserves the total number of fermions of each fermion type $f$. The conserved Abelian symmetries are manifest from the couplings of the form $\hat{\chi}_f^\dagger(r) \hat{\chi}_f(r+1)$ in Eq.~\eqref{eq: HI} directly in the interaction Hamiltonian:
global rotations of the form $\hat{\chi}_f(r) \to e^{i \phi} \hat{\chi}_f(r)$ leave the Hamiltonian invariant. These U(1) transformations consequently preserve the following global fermionic quantum numbers:
\begin{equation}
    \sum_{r=1}^{N} \nu_{\obar}(r) ~,~ \sum_{r=1}^{N} \num(r) ~,~ \sum_{r=1}^{N} \nu_{1}(r).
    \label{eq: global conserved charges LSH}
\end{equation}
The global symmetries of the Hamiltonian as characterized above are generated by the associated total fermion number operators, $\sum_r \hat{\nu}_f(r)$, which all mutually commute with each other and independently generate their own Abelian symmetry.

More insightful linear combinations of these three global U(1) conserved charges in Eq.~\eqref{eq: global conserved charges LSH} are given by 
\begin{align}
    \mathcal{F} &= \sum_{r=1}^{N} \big( \nu_{\obar}(r) +  \num(r) + \nu_{1}(r)\big) ,
    \label{eq: conserved total fermi number}\\
    \Delta\mathcal{P} &= \sum_{r=1}^{N} \big(\nu_1(r)-\nu_0 (r) \big) ,
    \label{eq: conserved total P}\\
    \Delta\mathcal{Q} &= \sum_{r=1}^{N} \big(\num(r)-\nu_{\obar} (r) \big),
    \label{eq: conserved total Q}
\end{align}
where $\mathcal{F}$ is the total fermion number that can take integer values from 0 to $3N$.
The charges $\Delta\mathcal{P}$ and $\Delta\mathcal{Q}$ denote the net bosonic flux of $P$-type and $Q$-type that the system sources or sinks when traversed from left to right.
Thus, for given (background) gauge fluxes fixed at the first lattice site, $(P(\obar,1),Q(\obar,1)) = (\mathcal{P}_0,\mathcal{Q}_0)$, the total gauge flux at the last site is then given by $(P(1,N),Q(1,N))=(\mathcal{P}_0+\Delta\mathcal{P}, \mathcal{Q}_0+\Delta\mathcal{Q})$.
Equations~\eqref{eq: P1Q1} and~\eqref{eq: P2Q2} along with the Gauss's law conditions in Eq.~\eqref{eq: AGL} imply that each lattice site can source and/or sink up to one unit of gauge flux of either type, indicating that $\Delta \mathcal{P}$ and $\Delta\mathcal{Q}$ can both take integer values from $-N$ to $N$.
We will elaborate on this later in Sec.~\ref{subsec: LSH-basis} in conjunction with a pictorial illustration in Fig.~\ref{fig: pict-LSH}.

An important subspace of the Hilbert space is the space that contains the strong-coupling vacuum, i.e., the ground state of the theory in the limit of $x\to 0$.
This state is characterized as the state with no bosonic excitations at any site and with fermionic modes at the sites alternating between completely empty and completely full. 
Thus the strong-coupling vacuum belongs to the global symmetry sector characterized by
\begin{align}
    \sum_{r=1}^{N} \nui(r) = \sum_{r=1}^{N} \num(r) = \sum_{r=1}^{N} \nuo(r) = \frac{N}{2} ,
    \label{eq: strong coupling vacuum charges LSH}
\end{align}
or equivalently,
\begin{align}
    \mathcal{F} = \frac{3N}{2}, \quad (\Delta\mathcal{P},\Delta\mathcal{Q}) = (0,0).
    \label{eq: strong coupling vacuum charges FPQ}
\end{align}
This is the sector that contains all states in the Hilbert space that can be reached from the strong coupling vacuum via the dynamics governed by the Hamiltonian in Eq.~\eqref{eq: ham-terms}.
This sector is especially important for periodic boundary conditions where $(\Delta\mathcal{P},\Delta\mathcal{Q})=(0,0)$ necessarily.

\subsection{Numerical benchmarking of energy eigenstates
\label{subsec: result-numerics}
}
\begin{figure}[t]
    \centering
    \begin{subfigure}{0.4\textwidth}
        \captionsetup{justification=centering}
        \includegraphics[width=\textwidth]{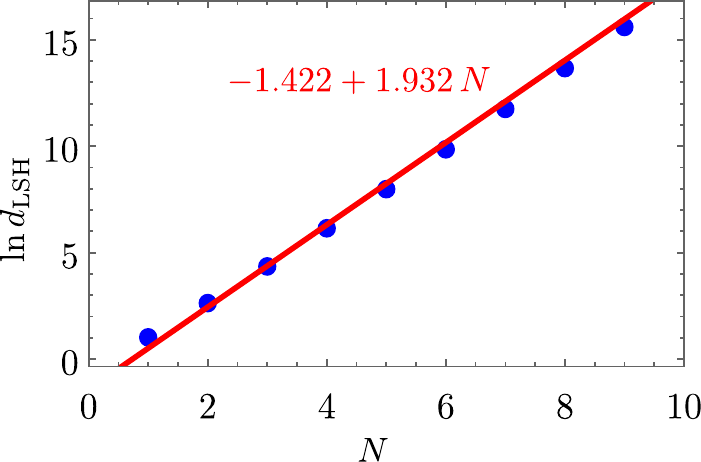}
        \caption{}
        \label{subfig: log dim vs N}
    \end{subfigure}
    \begin{subfigure}{0.584\textwidth}
        \captionsetup{justification=centering}
        \includegraphics[width=\textwidth]{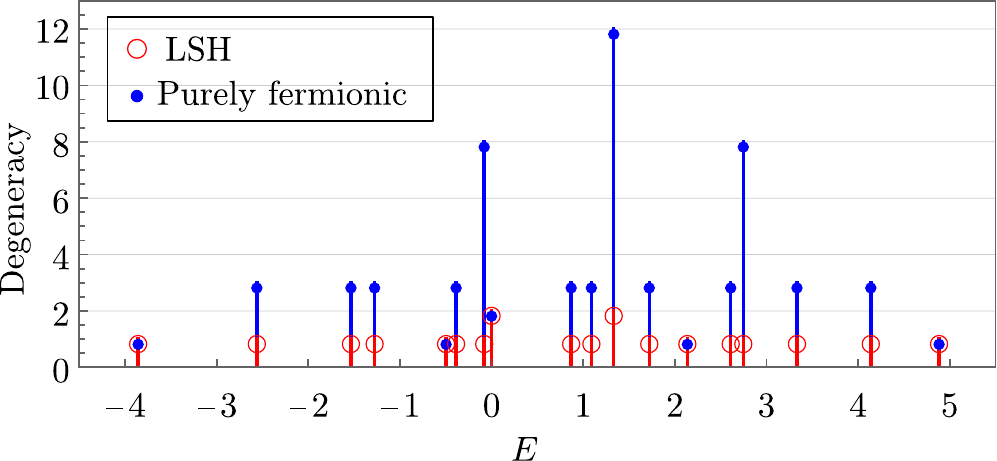}
        \caption{}
        \label{subfig: eigenavlues}
    \end{subfigure}
    \caption{(a) The dependence of the logarithm of the dimension of the physical Hilbert space within the LSH formulation, $d_{\rm LSH}$, is plotted for a range of lattice size values, $N$. For the purposes of this dimensionality comparison, the range of $N$ has been artificially extended to include odd values.   (b) Degeneracy in eigenvalues are plotted against their values for the Hamiltonian with $\mu = x = 1$ that is defined on a lattice of size $N=2$ with open boundary condition and zero incoming flux. Note that the eigenvalues are those of the dimensionless Hamiltonian in Eq.~\ref{eq: ham-terms}. The eigenvalues are calculated with the LSH framework (red) and the gauge fixed purely fermionic framework (blue). The eigenvalues in both formulations match exactly with each other up to machine precision, however the degeneracy factor is different as discussed in Sec.~\ref{subsec: result-numerics}.\label{fig: spectrum}}
\end{figure}
The dynamics of the (1+1)-dimensional SU(3) lattice gauge theory with staggered quarks is completely captured by the LSH Hilbert space and the LSH Hamiltonian introduced in the previous subsections.
The derivation of the LSH formulation from the SU(3) Schwinger boson formulation will come in Sec.~\ref{sec: LSH-framework}, but first its validity will be confirmed by comparing its energy eigenvalues with another formulation that was recently considered in Refs.~\cite{Farrell:2022wyt,Atas:2022dqm}.
This formulation, which we refer to as ``purely fermionic,'' is the result of taking the Kogut-Susskind formulation of the model with open boundary conditions and completely integrating out the gauge field (leaving only the colored fermionic field degrees of freedom);
a brief derivation is outlined in Appendix~\ref{app: purely fermionic formulation}.
Below, we discuss the results of this numerical benchmarking against the purely fermionic formulation.

The Hilbert space of this model is spanned by the LSH occupation-number basis states which, are tensor product states of on-site LSH basis states in Eq.~\eqref{eq: LSH-state} across the lattice subjected to the Abelian Gauss's law constraints given in Eqs.~\eqref{eq: AGL} on all links. Note that Eqs.~\eqref{eq: P1Q1} and~\eqref{eq: P2Q2} imply that the matter content at each site can increase or decrease the $P$-type or $Q$-type gauge fluxes by at most 1 unit when traversing a site. As a result, the physical Hilbert space of an $N$-site lattice with open boundary condition and given gauge fluxes fixed at either one of the boundaries (e.g., the first lattice site) is finite-dimensional. That said, the dimension of the physical Hilbert space still grows exponentially with the lattice size. This is illustrated empirically in the plot in Fig.~\ref{subfig: log dim vs N} for the case of $N\leq10$ lattices with zero background flux.
Here, the LSH physical Hilbert space dimension, $d_{\rm LSH}$, is observed to scale with $N$ as 
\begin{equation}
    d_{\rm LSH}\approx \exp (-1.422+1.932N).
    \label{eq: LSH Hilbert space dim fit}
\end{equation}

The Hilbert space in the purely fermionic formulation for the same background flux is spanned by a basis that is characterized by the fermionic occupation number for each color index at each lattice site (see Eq.~\ref{eq: Fermionic basis state} in Appendix~\ref{app: purely fermionic formulation}). The dimension of this Hilbert space is then $8^N$ that is given by the number of all possible colored fermionic occupation states, which implies that the dimension of the Hilbert space in purely fermionic formulation is always greater than the dimension of the LSH Hilbert space. To compare, the ratio of the Hilbert space dimension for the fermionic formulation to that of the LSH formulation for lattices in Fig.~\ref{subfig: log dim vs N} is given by
\begin{equation*}
    \ln{d_F/d_{LSH}} = N(\ln 8 -1.932) +1.422 \Rightarrow d_F/d_{LSH}\propto \exp{0.148N}.
\end{equation*}

Despite their different dimensions, the energy eigenvalues in both formulations are the same. This is demonstrated in the plot in Fig.~\ref{subfig: eigenavlues} that compares the energy spectra of the Hamiltonian with $\mu=x=1$ and zero flux to the left of site $r=1$ for an $N=2$ lattice, calculated in both the LSH and purely fermionic formulations. The energy eigenvalues match up to machine precision, providing a numerical benchmark for the LSH Hamiltonian. 

The excess of states in the Hilbert space of the purely fermionic formulation, seen in Fig.~\ref{subfig: eigenavlues}, is attributed to the fact that the basis states belong to irreps of the global SU(3) symmetry as discussed in Appendix~\ref{app: purely fermionic formulation}.
For the simplified case of zero boundary flux at the leftmost site, the purely fermionic states in each $\mathcal{F}$ sector are multiplets identified by the remaining two global charges $(\Delta\mathcal{P},\Delta\mathcal{Q})$ which in this case are non-negative integers and characterize the global SU(3) irrep.
In the more general case with a nonzero left-boundary flux in the $(\mathcal{P}_0,\mathcal{Q}_0)$ irrep, the multiplets are now identified by two non-negative integers $(\mathcal{P}_f,\mathcal{Q}_f)$  such that $(\mathcal{P}_f,\mathcal{Q}_f) =(\mathcal{P}_0+\Delta\mathcal{P}, \mathcal{Q}_0+\Delta\mathcal{Q})$.
Each state in the set of degenerate eigenstates with $(\mathcal{P}_f,\mathcal{Q}_f)$ charges is then further distinguished by its isospin and hypercharge quantum numbers, leading to a multiplicity $d({\mathcal{P}_f,\mathcal{Q}_f})$ given by the dimension of the irrep $(\mathcal{P}_f,\mathcal{Q}_f)$:
\begin{equation}
    d(\mathcal{P}_f, \mathcal{Q}_f) = \frac{1}{2}(\mathcal{P}_f+1)(\mathcal{Q}_f+1)(\mathcal{P}_f+\mathcal{Q}_f+2) .
    \label{eq: dim of irrep}
\end{equation}
On the other hand, the same global symmetry sectors in LSH formulation are now described by U(1) charges $\mathcal{F}$ and $(\Delta\mathcal{P},\Delta\mathcal{Q})$ as mention in Sec.~\ref{subsec: super-selection}. 
This formulation has no degeneracy associated with SU(3) gauge transformations because the states are all fundamentally constructed in terms of SU(3)-singlet operators.
As a result, for every LSH eigenstate with a $(\mathcal{P}_f,\mathcal{Q}_f)$ global charge, there are $d(\mathcal{P}_f,\mathcal{Q}_f)$ degenerate purely fermionic states. This is further demonstrated for the case of eigenvalues in Fig.~\ref{subfig: eigenavlues} in Table~\ref{tab: Eigenvalues} in Appendix~\ref{app: purely fermionic formulation}.

\section{Kogut-Susskind framework and its reformulation in terms of prepotentials
\label{sec: methods1}
}
\noindent
In Sec.~\ref{sec: results}, we discussed the LSH quantum numbers, construction of LSH Hilbert space through the Abelian Gauss's law and the Hamiltonian in terms ladder and number operators of LSH degrees of freedom. There still remains to show the derivation of this framework starting from the prepotential Hamiltonian formulation of lattice gauge theory. In this section we review the Kogut-Susskind Hamiltonian for an 1+1 dimensional SU(3) lattice gauge theory in Sec.~\ref{subsec: KS-framework} and the equivalent prepotential formulation for describing the gauge bosons in Sec.~\ref{subsec: prepotential}. Then, in Sec.~\ref{subsec: prepotential-with-matter}, we couple the prepotential degrees of freedom to matter fields in the form of one flavor of staggered fermions.

\subsection{Canonical framework for Hamiltonian lattice gauge theory
\label{subsec: KS-framework}
}

\begin{figure}
    \centering
    \begin{subfigure}{\textwidth}
        \captionsetup{justification=centering}
        \includegraphics[width=\textwidth]{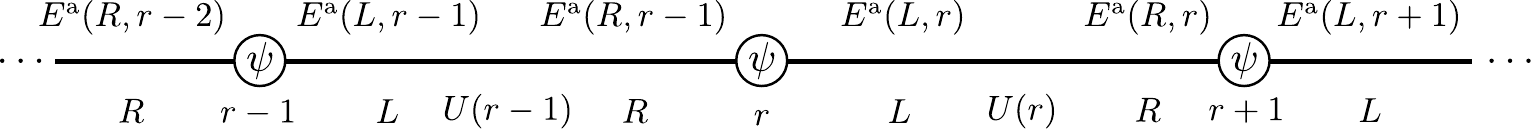}
        \caption{Kogut-Susskind variables}
        \label{subfig: KS variables}
    \end{subfigure}
    \par\bigskip
    \begin{subfigure}{\textwidth}
        \captionsetup{justification=centering}
        \includegraphics[width=\textwidth]{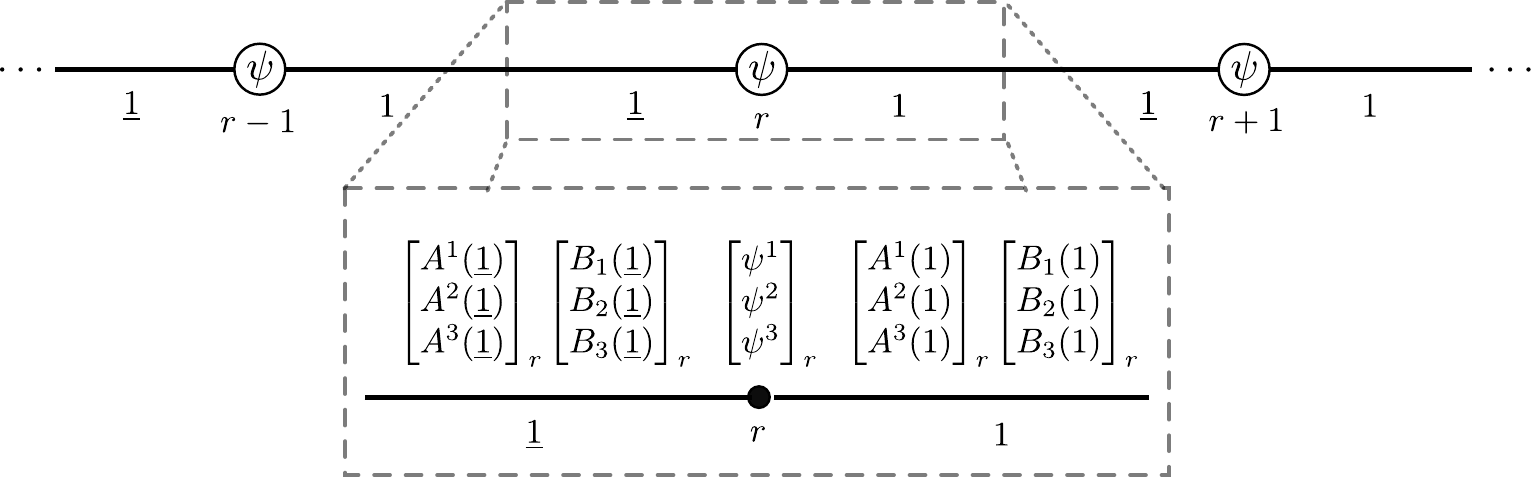}
        \caption{Prepotential variables.}
        \label{subfig: prepotential variables}
    \end{subfigure}
\caption{(a) Three adjacent lattice sites at $r-1$, $r$, and $r+1$, and their links with Kogut-Susskind degrees of freedom. The left and right ends of a link are denoted by $L$ and $R$, respectively. The chromoelectric field, $E^{\rm a}$ is labeled by a side, $L$ or $R$, denoting the end of the link at which it resides, and a position argument indicating the corresponding link. The link operator, $U$, is defined on each link as denoted by its position argument, and the staggered matter, $\psi$, is defined at each lattice sites denoted by a circle in this figure. (b) The prepotential degrees of freedom for the same lattice as in Fig.~\ref{subfig: KS variables}. 
The prepotential formulation of the same theory introduces four independent sets of SU(3) irreducible Schwinger bosons for each site $r$: $A^\alpha(1,r)$ and $B_\alpha(1,r)$ residing on the $L$ side of link $r$, and $A^\alpha(\obar,r)$ and $B_\alpha(\obar,r)$ residing on the $R$ side of link $r-1$, such that degrees of freedom are essentially site-localized.
In tandem with the emphasis on site-local degrees of freedom, the conventional $L$ and $R$ labels for left and right ends of links are replaced by $1$ and $\obar$, respectively.
Additionally, for brevity in this figure, we have used the shorthand notation similar to Eq.~\eqref{eq: HI} for expressing the position labels of $A$, $B$ and $\psi$.
}
\end{figure}

The Kogut-Susskind Hamiltonian for the 1+1 dimensional SU(3) lattice gauge theory with staggered quarks is defined on a spatial lattice with $N$ staggered lattice sites (or $N/2$ ``physical'' sites) and continuous time. We refer to a lattice site by its position $r=1,2,\cdots,N$ and to the link joining sites $r$ and $r+1$ by `$r$' as well. The gauge degrees of freedom in this formulation are described by the canonically conjugate variables on each link, namely, the chromoelectric fields and the holonomy or link operators that remain after choosing the temporal gauge. The chromoelectric fields reside at the left, $L$, and the right, $R$, ends of each link $r$, and they are denoted by $E^{\rm a}(L,r)$ and $E^{\rm a}(R,r)$, respectively, where `a' is the adjoint index that takes values from ${\rm a}=1,\cdots,\,8$. Meanwhile, the link operators are unitary $3\times 3$ matrices defined on each link $r$, with components denoted by $U^\alpha{}_\beta(r)$, where $\alpha$, $\beta =$ 1, 2, 3 are the color indices. The electric fields and link operators are illustrated together in Fig.~\ref{subfig: KS variables}. The canonical commutation relations are given by
\begin{eqnarray}
  [E^{\rm a}(L/R,r),E^{\rm b}(L/R,r')] &=& \delta_{rr'} \sum_{{\rm c}=1}^8 i\,f^{\rm abc}\,E^{\rm c}(L/R,r),
  \label{eq: commutation-ELa-Elb}\\
  \left[E^{\rm a}(L,r),U(r')\right] &=& -\delta_{rr'}\,T^{\rm a} U(r)\,,\quad [E^{\rm a}(R,r),U(r')]= \delta_{rr'}\,U(r)T^{{\rm a}},
  \label{eq: commutation-ELa-U}\\
  \left[U^\alpha{}_\beta(r),U^\gamma{}_\eta(r')\right] &=& [U^\alpha{}_\beta(r),U^{\dagger\gamma}{}_\eta(r')]=0,
  \label{eq: commutation-U-U}
\end{eqnarray}
where $f^{\rm abc}$ are the structure constants for SU(3), $T^{\rm a}=\frac{\lambda^{\rm a}}{2}$ with $\lambda^{\rm a}$ being the Gell-Mann matrices, and $L/R$ indicates that the relation holds individually for both $L$ and $R$ sides of the link. The left and right chromoelectric fields on a link are connected by parallel transport, that is they are related to each other via the relation
\begin{equation*}
    E^{\rm a}(R,r)T^{\rm a}=-U^\dagger(r)E^{\rm b}(L,r)T^{\rm b} U(r).
\end{equation*}
As a consequence, the quadratic Casimirs on either side, $E^2(r)$, must be equal:
\begin{equation}
    E^2(r)\equiv \sum_{\rm a} E^{\rm a}(L,r)E^{\rm a}(L,r)=\sum_{\rm a} E^{\rm a}(R,r)E^{\rm a}(R,r).
    \label{eq: electric field casimir constraint}
\end{equation}
The one-flavor fermionic matter is expressed by a staggered fermionic field at each lattice site $r$, $\psi_\alpha(r)$, where $\alpha$ is again the color index. It obeys the fermionic anticommutation relations given by
\begin{equation}
    \{\psi_\alpha^\dagger(r),\psi^\dagger_\beta(r')\}=\{\psi^\alpha(r),\psi^\beta(r')\}=0,
    \qquad \{\psi^\alpha(r),\psi^\dagger_\beta(r')\}=\delta^{\alpha}_\beta\,\delta_{rr'}. 
    \label{eq: ferm_anticomm}
\end{equation}
The Kogut-Susskind Hamiltonian for an $N$-site lattice is then given by
\begin{eqnarray}
  H&=& H_M+H_E+ H_{I} \nonumber\\ 
  &=& \mu \sum_{r=1}^{N} (-1)^r \psi^\dagger_\alpha(r) \psi^\alpha (r) + \sum_{r=1}^{N'}  E^2(r) + x \sum_{r=1}^{N'} \left[\psi^\dagger_\alpha(r)\, U^\alpha{}_{\beta}(r)\, \psi^\beta(r+1) + {\rm H.c.} \right] ,
  \label{eq: KS-ham}
\end{eqnarray}
where $H_M$ is the matter self-energy, $H_E$ is the chromoelectric energy, and $H_{I}$ is the gauge-matter interaction. Here, repeated indices are summed over, and $N'=N-1$ $(N)$ for open (periodic) boundary conditions. The dimensionless couplings $\mu$ and $x$ are related to the fermion mass and the coupling constant, respectively; see Ref.~\cite{Hamer:1981yq} for details.

The Hamiltonian in Eq.~\eqref{eq: KS-ham} is accompanied by a constraint structure, known as the
Gauss's law constraints, given by
\begin{equation}
  G^{\rm a}(r)|\Psi\rangle=0 \quad \forall\, {\rm a}\,,r.
  \label{eq: gauss-law}
\end{equation}
where $G^{\rm a}(r)$ are the full generators of the local gauge transformations:
\begin{equation}
  G^{\rm a}(r)= E^{\rm a}(L,r)+E^{\rm a}(R,r-1)+ \psi^\dagger_\alpha(r)\, \left(T^{\rm a}\right)^\alpha{}_\beta\, \psi^\beta(r).
  \label{eq: gauss-op}
\end{equation}
The gauge invariance of Hamiltonian means that it commutes with $G^{\rm a}(r)$ for all ${\rm a}$, $r$. The Hamiltonian generates the dynamics between states of the Hilbert space, $|\Psi\rangle$, that satisfy the Gauss's laws.

The convenient choice of basis for Kogut-Susskind formulation is the strong coupling basis, where the gauge-invariant ground state is defined as having zero electric flux, no particles, and no antiparticles. The gauge-matter interaction Hamiltonian $H_{I}$, when applied to this state, builds up the physical Hilbert space of the theory. However, the physical Hilbert space is only a really small subspace of the full gauge theory Hilbert space that is created by applying individual link operators on the strong coupling vacuum. The gauge-invariant Hilbert space contains non-local Wilson loops, non-local strings, and hadrons. Such a space of states, conventionally known as the ``loop space,'' albeit gauge invariant, is spanned by an over-complete and non-local basis that is not that useful for practical computation \cite{Mathur:2007nu}. However, the reformulation of the theory in terms of prepotentials leads to a local and complete basis for the physical Hilbert space of the theory \cite{Mathur:2004kr}.

\subsection{Prepotential formulation
\label{subsec: prepotential}
}
\noindent
The prepotential formulation of gauge theories, that started developing more than a decade ago for SU(2), gives a reformulation of the same Kogut-Susskind theory allowing one to construct a local and complete gauge invariant basis. Generalization of the prepotential formulation for SU(3) as well as for arbitrary SU(N) is an involved process~\cite{Anishetty:2009ai, Anishetty:2009nh, Raychowdhury:2013rwa}. In this subsection, we will briefly summarize the key results for SU(3).

Prepotentials are the Schwinger boson (or harmonic oscillator) operators, in the fundamental representation of the gauge group. For the SU(2) gauge group, which is a rank one group with a two dimensional fundamental representation, the set of prepotential operators consists of one Schwinger boson doublet at the $L$ and $R$ ends of each link. Similarly, the prepotential formulation of the SU(3) gauge group, which is a rank-two group with a three-dimensional fundamental representation, requires two independent Schwinger boson triplets at both ends of each link. They are denoted by $a^\dagger_\alpha(L,r)$ and $b^{\dagger\alpha}(L,r)$ for Schwinger bosons on the $L$ end of link $r$, and $a^\dagger_\alpha(R,r)$ and $b^{\dagger\alpha}(R,r)$ for Schwinger bosons on the $R$ end of the link $r-1$. Here, $\alpha$ and $\beta$ are the color indices that take integer values from 1 to 3. Note that, whereas bosonic variables of the Kogut-Susskind formulation are identified by their links, the Schwinger boson definitions are focused on sites. The SU(3) Schwinger bosons obey the ordinary bosonic commutation relations:
\begin{align}
    \big[a^\alpha(s,r),a^{\alpha'}(s',r')\big]&=\big[b_\alpha(s,r),b_{\alpha'}(s',r')\big]=0,
    \label{eq: SU3 schwinger boson aa bb commutation}\\
    \big[a^\alpha(s,r),b_{\alpha'}(s',r')\big]&=\big[a^\alpha(s,r),b^{\dagger\alpha'}(s',r')\big]=0,
    \label{eq: SU3 schwinger boson ab commutation}\\
    \big[a^\alpha(s,r),a^\dagger_{\alpha'}(s',r')\big]&=\big[b_{\alpha'}(s,r),b^{\dagger\alpha}(s',r')\big]=\delta_\alpha^{\alpha'}\delta_{ss'}\delta_{rr'}
    \label{eq: SU3 schwinger boson non-zero commutation},
\end{align}
where $s$ and $s'$ can each take values $L$ or $R$.

Following the Schwinger boson construction for SU(3) generators, the left and right chromoelectric fields on a link could be defined in terms of the prepotentials as
\begin{subequations}
\label{eq: E in terms of a and b}
\begin{align}
    E^{\rm a}(L,r)&= a^\dagger_\alpha(L,r)\,\big(T^{\rm a}\big)^\alpha{}_\beta \, a^\beta(L,r) - b^{\dagger\alpha}(L,r)\,\big(T^{*\rm a}\big)_\alpha{}^\beta\, b_\beta(L,r) , \\
    E^{\rm a}(R,r-1)&= a^\dagger_\alpha(R,r)\,\big(T^{\rm a}\big)^\alpha{}_\beta \, a^\beta(R,r) - b^{\dagger\alpha}(R,r)\,\big(T^{*\rm a}\big)_\alpha{}^\beta\, b_\beta(R,r) ,
\end{align}
\end{subequations}
with $T^{\rm a}$ and the $L/R$ notation being the same as in Eqs.~\eqref{eq: commutation-ELa-Elb} and~\eqref{eq: commutation-ELa-U}. It was shown in \cite{Anishetty:2009ai,Anishetty:2009nh} that this particular set of Schwinger bosons are not suitable for the construction of the gauge theory Hilbert space. To appreciate this fact, it is instructive to look at a simpler example of SU(2) gauge bosons.

In the case of the SU(2) gauge group, only one doublet of Schwinger bosons, $a^\dagger_\alpha$ with $\alpha=1,2$ is needed to represent the SU(2) irreps, as mentioned before. An SU(2) irrep corresponding to angular momentum $j$ is constructed in terms of these Schwinger bosons as
\begin{equation}
    |j\rangle_{\vec{\alpha}} = \mathcal N a^\dagger_{\alpha_1}\ldots a^\dagger_{\alpha_{2j}}|0\rangle,
    \label{eq: SU2-j-state}
\end{equation}
where $\vec{\alpha} = (\alpha_1,\cdots,\alpha_{2j})$, $\mathcal N$ is the normalization factor, and the $j=0$ state, $|0\rangle$, is defined by the condition $a_1 |0\rangle=a_2 |0\rangle=0$. The indices $\alpha_{i}$ for $i=1,2, \dots, 2j$ can take any value between 1 and 2, and their values determine the azimuthal quantum number $m$ of that irrep. (We have left the azimuthal quantum number in Eq.~\eqref{eq: SU2-j-state} implicit for brevity.) A similar monomial of the Schwinger bosons $a^\dagger_{\alpha}$ and $b^{\dagger\beta}$ will likewise create representations of SU(3), however they are generally not irreducible~\cite{mukunda1965tensor,Chaturvedi:2002si}. The simplest case of the $(1,1)$ irrep of SU(3) is naively constructed as $a^\dagger_{\alpha} b^{\dagger\beta}|0,0\rangle$, however its proper construction is instead
\begin{equation}
   |1,1\rangle_\alpha^\beta= a^\dagger_{\alpha} b^{\dagger\beta}|0,0\rangle - \frac{\delta_{\alpha}^{\beta}}{3} (a^\dagger \cdot b^\dagger)|0,0\rangle,  \label{eq: SU3-11-irrep}
\end{equation}
such that the irrep is traceless, which is a fundamental property of any irrep. Above, $\alpha\,,\beta=1, 2, 3$ are the color indices,  $a^\dagger\cdot b^\dagger \equiv \sum_{\gamma=1}^3 a^\dagger_\gamma b^{\dagger\gamma}$, and the $(0,0)$ irrep state, $|0,0\rangle$, satisfies $a^\alpha|0,0\rangle = b_{\alpha}|0,0\rangle=0$ for all values of $\alpha$. For a general irrep $(P,Q)$, one has to extract out all the traces from the monomial state $a^\dagger_{\alpha_1}\cdots a^\dagger_{\alpha_P} b^{\dagger\beta_1}\cdots b^{\dagger\beta_Q}|0,0\rangle $ to satisfy the tracelessness condition (see, for example, equation (35) of Ref.~\cite{Mathur:2000sv}). Such a traceless construction is cumbersome to use, and on top of that there is also a multiplicity problem because $(a^\dagger\cdot b^{\dagger})^{\rho}\displaystyle|P,Q\rangle_{\vec{\alpha}}^{\vec{\beta}}$ for any positive integer $\rho$ transforms in the same way under SU(3) as the $(P,Q)$ irrep \cite{Chaturvedi:2002si}.
A novel solution that uses the naive Schwinger bosons discussed thus far to construct a set of modified Schwinger bosons, called irreducible Schwinger bosons (ISBs), has previously been proposed in the Ref.~\cite{Anishetty:2009ai}, providing a monomial construction of traceless SU(3) irreps that solves the multiplicity problem.

The ISBs are constructed in terms of the naive SU(3) Schwinger bosons as
\begin{eqnarray}
    A^{\dagger}_{\alpha}&\equiv& a^{\dagger}_{\alpha}-\frac{1}{\hat{N}_a+ \hat{N}_b+1}(a^\dagger\cdot b^\dagger)b_{\alpha},
    \label{eq: Adagg-def} \\
    B^{\dagger\alpha}&\equiv& b^{\dagger\alpha}-\frac{1}{\hat{N}_a+\hat{N}_b+1}(a^\dagger\cdot b^\dagger)a^{\alpha}.
    \label{eq: Bdagg-def}
\end{eqnarray}
Above,
\begin{equation}
    \hat{N}_a \equiv a^{\dagger}\cdot a = \sum_{\alpha=1}^{3} a^{\dagger}_\alpha\,a^{\alpha} \quad\text{and}\quad \hat{N}_b \equiv b^{\dagger}\cdot b = \sum_{\beta=1}^{3} b^{\dagger\beta}\,b_{\beta} 
    \label{eq: prepotential number operators}
\end{equation}
are the total number operators for $a$-type and $b$-type Schwinger bosons, respectively. Using ISBs, the state in Eq.~\eqref{eq: SU3-11-irrep} may now be expressed as a monomial state: $|1,1\rangle^\beta_\alpha= A^\dagger_{\alpha} B^{\dagger\beta}|0,0\rangle$. In general, the monomial states constructed out of the ISBs in Eq.~\eqref{eq: Adagg-def} and~\eqref{eq: Bdagg-def} as
\begin{equation}
    |P,Q\rangle_{\vec{\alpha}}^{\vec{\beta}}= \mathcal{N} A^\dagger_{\alpha_1}\ldots A^\dagger_{\alpha_P} B^{\dagger\beta_1}\ldots B^{\dagger\beta_Q}|0\rangle
    \label{eq: SU3-irrep}
\end{equation}
are indeed the traceless SU(3) irreps. Here $\mathcal N$ corresponds to the normalization factor, the indices $\alpha_{i}$ with $i=1, \cdots, P$ and  $\beta_{j}$ with $j=1, \cdots, Q$ can take integer values between 1 to 3, and $\alpha_{i}$ and $\beta_{j}$ determine the isospin and hypercharge of the irrep state~\cite{mukunda1965tensor,Chaturvedi:2002si}. The monomial states constructed in Eq.~\eqref{eq: SU3-irrep} also solve the multiplicity problem as the states satisfy 
\begin{equation}
    A^\dagger\cdot B^\dagger\, |P,Q\rangle_{\vec{\alpha}}^{\vec{\beta}} =0  \quad\text{and}\quad A\cdot B\, |P,Q\rangle_{\vec{\alpha}}^{\vec{\beta}} =0,
    \label{eq: ISB multiplicity constraint}
\end{equation}
which ensures that they are restricted to the $\rho=0$ subspace or equivalently the kernel of the operator $a\cdot b$, see Ref.~\cite{Anishetty:2009nh} for details. This implies that the operators $A^\dagger\cdot B^\dagger$ and $A\cdot B$ are effectively the null operators within the Hilbert space spanned by states in Eq.~\eqref{eq: SU3-irrep}. This leads to the following modified commutation relations for ISBs:
\begin{eqnarray}
    &&[A^\alpha, A^\dagger_\beta] \simeq \left( \delta^\alpha_{\beta}-\frac{1}{\hat{N}_a+\hat{N}_b+2}B^{\dagger\alpha}B_\beta \right),
    \label{eq: AAdagg-commutator}\\
    &&[B_\alpha, B^{\dagger\beta}] \simeq  \left( \delta_\alpha^{\beta}-\frac{1}{\hat{N}_a+\hat{N}_b+2}A^{\dagger}_{\alpha}A^\beta \right),
    \label{eq: BBdagg-commutator}\\
    && [A^\alpha,B^{\dagger\beta}] \simeq   -\frac{1}{\hat{N}_a+\hat{N}_b+2}B^{\dagger\alpha}A^\beta, 
    \label{eq: ABdagg-commutator}\\
    && [B_\alpha,A^\dagger_\beta] \simeq   -\frac{1}{\hat{N}_a+\hat{N}_b+2}A^{\dagger}_\alpha B_\beta,
    \label{eq: AdaggB-commutator}
\end{eqnarray}
along with
\begin{eqnarray}
    [ A^\dagger_\alpha, A^\dagger_{\beta}]=[ A^\alpha, A^{\beta}]=[B_\alpha,B_\beta]=[ B^{\dagger\alpha}, B^{\dagger\beta}]=[A^\alpha,B_\beta]=[A^\dagger_\alpha, B^{\dagger\beta} ]=0,
    \label{eq: zero-commutators}
\end{eqnarray}
where $\simeq$ indicates that the above set of commutation relations are valid within the vector subspace spanned by SU(3) irreps defined in Eq.~\eqref{eq: SU3-irrep}. Note that, in the same subspace, the number operators for $A$-type and $B-$type ISBs satisfy
\begin{equation}
    \hat{N}_A \equiv A^\dagger\cdot A \simeq \hat{N}_a \quad \text{and} \quad \hat{N}_B \equiv B^\dagger\cdot B \simeq \hat{N}_b\,.
    \label{eq: NA Na and NB Nb equivalence}
\end{equation}

One can then proceed towards the prepotential framework for SU(3) lattice gauge theory as a straightforward extension of the SU(2) case using the SU(3) ISBs given in Eqs.~\eqref{eq: Adagg-def} and~\eqref{eq: Bdagg-def}. Sets of ISB prepotential operators  $A^\dagger_\alpha(L,r)$ and $B^{\dagger\alpha}(L,r)$ are introduced for the $L$ end of link $r$, and $A^\dagger_\alpha(R,r)$ and $B^{\dagger\alpha}(R,r)$ are introduced for the $R$ end of link $r-1$. These ISBs with different arguments commute, and they obey the commutation relations in Eqs.~\eqref{eq: AAdagg-commutator}-~\eqref{eq: zero-commutators} only if they have the same direction and position arguments. 
The electric fields in terms of ISBs follow the same format that was used in the naive definitions of Eqs.~\eqref{eq: E in terms of a and b}:
\begin{subequations}
\begin{align}
    E^{\rm a}(L,r)= A^\dagger_\alpha(L,r)\,\big(T^{\rm a}\big)^\alpha{}_\beta \, A^\beta(L,r) - B^{\dagger\alpha}(L,r)\big(T^{*\rm a}\big)_\alpha{}^\beta B_\beta(L,r). \\
    E^{\rm a}(R,r-1)= A^\dagger_\alpha(R,r)\,\big(T^{\rm a}\big)^\alpha{}_\beta \, A^\beta(R,r) - B^{\dagger\alpha}(R,r)\big(T^{*\rm a}\big)_\alpha{}^\beta B_\beta(R,r).
\end{align}
\label{eq: E in terms of A and B}
\end{subequations}
In order for this redefinition to satisfy Eq.~\eqref{eq: electric field casimir constraint}, the prepotential number operators \eqref{eq: NA Na and NB Nb equivalence} defined at each end of the link must be related to each other by either $(\hat{N}_A(L,r) - \hat{N}_A(R,r+1)) \ket{\Psi} = (\hat{N}_B(L,r) - \hat{N}_B(R,r+1)) \ket{\Psi} = 0$, or
\begin{align}
    (\hat{N}_A(L,r) - \hat{N}_B(R,r+1)) \ket{\Psi} =
    (\hat{N}_B(L,r) - \hat{N}_A(R,r+1)) \ket{\Psi} = 0
    \label{eq: AGL in ISB}
\end{align}
for any state $\ket{\Psi}$ belonging to the Hilbert space created by ISBs.
However, the fact that the gauge theory Hilbert space is built up by the action of link operators $U^\alpha{}_\beta$, transforming as in Eq.~\eqref{eq: commutation-ELa-U} is only consistent with the second choice of the constraint. Hence, the ISB construction of the link operator is defined to satisfy
\begin{align}
    [ U^\alpha{}_\beta(r) , \hat{N}_A(L,r) - \hat{N}_B(R,r+1)] = [ U^\alpha{}_\beta(r) , \hat{N}_B(L,r) - \hat{N}_A(R,r+1)] &= 0 .
    \label{eq: link operator AGL}
\end{align}
In the prepotential formulation of gauge theories, the constraints in Eq.~\eqref{eq: AGL in ISB} are known as the Abelian Gauss's law constraints of the theory, and they are responsible for the non-locality of the gauge-invariant (or physical) degrees of freedom of the theory. 

The link operator satisfying Eq.~\eqref{eq: link operator AGL} can be expressed in terms of the prepotential bosons as~\cite{Anishetty:2009nh}
\begin{align}
     U^\alpha{}_{\beta}(r)&= B^{\dagger\alpha}(L,r) \,\eta(r) \, A^{\dagger}_{\beta}(R,r+1) + A^{\alpha}(L,r)\, \theta(r)\, B_{\beta}(R,r+1)\nonumber\\
     &\quad + \bigl(A^{\dagger}(L,r) \wedge B(L,r)\bigl)^\alpha \, \delta(r) \, \bigl(B^{\dagger}(R,r+1)\wedge A(R,r+1)\bigr)_{\beta},
    \label{eq: link operator in prepotential}
\end{align}
where $(A^\dagger\wedge B)^\alpha \equiv \epsilon^{\alpha\gamma\delta}A^\dagger_{\gamma}B_{\delta}$ and $(B^\dagger\wedge A)_\beta \equiv \epsilon_{\beta\gamma\delta}B^{\dagger\gamma}A^{\delta}$, and the coefficients $\eta(r)$, $\theta(r)$, $\delta(r)$ are fixed by the unitarity and unit-determinant conditions on $U(r)$ given by $U^\dagger(r) U(r)= \mathds{1}_{3\times 3}$ and $\det U(r)=\mathds{1}$, respectively. It is important to note that just like in the prepotential formulation of the SU(2) gauge group, the link operator matrix $U(r)$ for the SU(3) gauge group can be written as a product of two SU(3) matrices defined in terms of the $(L,r)$ and $(R,r+1)$ prepotential operators \cite{Anishetty:2009ai,Anishetty:2009nh}. 

The motivation behind the prepotential formulation of lattice gauge theories is to be able to construct a local and minimal set of gauge-invariant operators along with an orthonormal Hilbert space for the pure Yang-Mills theory. For the one-dimensional lattice, the gauge theory becomes dynamical in the presence of matter fields. In the next subsection, we will discuss the coupling of prepotentials to staggered fermions.
\subsection{Irreducible prepotentials for SU(3) coupled to staggered matter
\label{subsec: prepotential-with-matter}
}
\begin{table}[t]
    \renewcommand{\arraystretch}{1.8}
    \centering
    \begin{tabular}{C{3 cm} | C{3 cm}}
        \hline
        Fundamental & Anti-fundamental \\
        $(1,0)$ or \textbf{3} & $(0,1)$ or \textbf{3*}\\
        \hline
        \hline
        $A^\dagger_{\alpha}(\obar,r)$ & 
        $A^\alpha(\obar,r)$\\
        $A^\dagger_{\alpha}(1,r)$ &
        $A^\alpha(1,r)$ \\
        $B_{\alpha}(\obar,r)$ & 
        $B^{\dagger\alpha}(\obar,r)$\\
        $B_{\alpha}(1,r)$ &
        $B^{\dagger\alpha}(1,r)$\\
        $\psi^\dagger_\alpha(r)$ &
        $\psi^{\alpha}(r)$ \\
        \hline
    \end{tabular}
    \caption{Prepotential variables and their irreps under the SU(3) gauge group at any site $r$. Equation~\eqref{eq: SU3-irrep} defines the irreps of $A^\dagger_\alpha$ and $B^{\dagger\alpha}$ ISBs. The choice of matter field irreps to be fundamental or anti-fundamental is arbitrary and it only affects the construction of gauge invariant singlets in Sec.~\ref{subsubsec: LSH-operators}; we have chosen the $\psi^\dagger(r)$ triplet to transform as the fundamental irrep under the SU(3) gauge transformations.\label{tab: prepotential irreps}}
\end{table}
To couple the prepotential operators with fermionic matter, we consider the Kogut-Susskind staggered fermionic matter field, $\psi^\alpha(r)$, at each lattice site $r$. It satisfies the anticommutation relations given in Eq.~\eqref{eq: ferm_anticomm} and transforms as a triplet under the SU(3) gauge group.
The matter field $\psi^\alpha(r)$ along with the prepotential operators, $A^\alpha(L,r)$, $B_{\alpha}(L,r)$, $A^\alpha(R,r)$ and $B_{\alpha}(R,r)$, forms the set of local prepotential variables that completely describe a one-dimensional SU(3) lattice gauge theory. 
Given the site-centric labeling of ISB degrees of freedom, and in anticipation of higher-dimensional generalizations, we relabel the $L$ and $R$ ends of links with $1$ and $\obar$, respectively. This choice of labels is not standard, so to familiarize readers with it, we contrast it against the conventional notation in Fig.~\ref{subfig: prepotential variables} by illustrating the prepotential variables with $1$ and $\obar$ labels next to Kogut-Susskind variables with $L$ and $R$ labels.

The prepotential variables transform as triplets under the SU(3) gauge group either as fundamental, $(1,0)$, or anti-fundamental, $(0,1)$, irreps. Equation~\eqref{eq: SU3-irrep} implies that oscillators of type $A^\dagger_\alpha$ transform as the $(1,0)$ irrep and the oscillators of type $B^{\dagger\alpha}$ transform as $(0,1)$ irrep, while their conjugates transform as $(0,1)$ and $(1,0)$ irreps, respectively. For the fermionic matter field, we have taken, without loss of generality,  $\psi^\dagger_\alpha(r)$ and $\psi^\alpha(r)$ to transform as $(1,0)$ and $(0,1)$ irreps, respectively. Thus, the fermionic matter at each lattice site is expressed in terms of fundamental, and not anti-fundamental, matter fields. The complete set of fundamental and anti-fundamental prepotential variables present at a lattice site are tabulated in Table~\ref{tab: prepotential irreps}. One can use the various fundamental and anti-fundamental fields to construct a complete set of local gauge singlets via symmetric or antisymmetric contractions with the Kronecker delta, $\delta_\alpha^\beta$, or Levi-Civita symbols, $\epsilon_{\alpha\beta\gamma}$ or $\epsilon^{\alpha\beta\gamma}$, respectively. We focus on a subset of these local singlets that acts as the on-site building blocks of the Kogut-Susskind Hamiltonian in Eq.~\eqref{eq: KS-ham} re-written in terms of prepotentials variables.

The mass Hamiltonian, $H_M$, in the staggered formulation as given in Eq.~\eqref{eq: KS-ham} is carried over unchanged to the prepotential formulation. The electric Hamiltonian, $H_E$, consisting of the quadratic Casimir operator defined at each end of a link is given by replacing chromoelectric fields in Eq.~\eqref{eq: KS-ham} with their corresponding prepotential definition given in Eqs.~\eqref{eq: E in terms of A and B}. This leads to
\begin{equation}
    H_E = \sum_{r=1}^{N'} H_E(r) = \sum_{r=1}^{N'} \frac{1}{3} \left( \hat{N}_A(1,r)^2 + \hat{N}_B(1,r)^2 + \hat{N}_A(1,r) \hat{N}_B(1,r) \right) + \hat{N}_A(1,r) + \hat{N}_B(1,r),
    \label{eq: HE in prepotential}
\end{equation}
Note that, because of the equality of Casimirs in Eq.~\eqref{eq: electric field casimir constraint}, it is arbitrary whether $H_E$ is defined in terms of $1$-side or $\obar$-side ISBs.

The dynamical part of the Hamiltonian is governed by the interaction between fermions and gauge fields denoted by $H_I$ in Eq.~\eqref{eq: KS-ham}. It contains local gauge singlet terms involving the gauge link operator $U(r)$, and the staggered matter field $\psi(r)$. The prepotential reformulation of $H_I$ is obtained by replacing $U(r)$ in the term $\psi^\dagger_\alpha(r)\, U^\alpha{}_{\beta}(r)\, \psi^\beta(r+1)$ with its expression in terms of prepotential variables given in Eq.~\eqref{eq: link operator in prepotential}. This leads to
\begin{align}
    \psi^\dagger_\alpha(r)\, U^\alpha{}_{\beta}(r)\, \psi^\beta(r+1) &=
    \left[\psi^\dagger_\alpha\, B^{\dagger\alpha}(1) \; \eta(1) \right]_r  \left[ \eta(\obar) \; \psi^\beta \, A^\dagger_\beta(\obar) \right]_{r+1} \nonumber\\
    & \quad + \,\left[\psi^\dagger_\alpha \, A^\alpha(1) \; \theta(1) \right]_r  \left[ \theta(\obar) \; \psi^\beta \, B_\beta(\obar) \right]_{r+1}\nonumber\\
    & \quad + \,\left[\psi^\dagger_\alpha \, (A^\dagger(1)\wedge B(1))^\alpha  \; \delta(1) \right]_r \left[ \delta(\obar) \; \psi^\beta \, (B^{\dagger}(\obar) \wedge A(\obar))_\beta\right]_{r+1},
    \label{eq: HI local term in prepotential}
\end{align}
where we have used the shorthand notation $[ \ ]_r$ for the position labels similar to Eq.~\eqref{eq: HI} for brevity. Above, each of the $\eta(r)$, $\theta(r)$ and $\delta(r)$ in Eq.~\eqref{eq: link operator in prepotential} have been factored into a product of two operators, one for each end of the link, as
\begin{equation}
    \eta(r) = \eta(1,r)\; \eta(\obar,r+1), \quad \theta(r) = \theta(1,r)\; \theta(\obar,r+1), \quad \text{and}\quad \delta(r) = \delta(1,r)\;\delta(\obar,r+1).
    \label{eq: eta theta delta factorization}
\end{equation}
The conditions $U^\dagger(r) U(r)= \mathds{1}_{3\times 3}$ and $\det U(r)=\mathds{1}$ on $U(r)$ in Eq.~\eqref{eq: link operator in prepotential} determine the form of these decomposed operators to be
\begin{align}
    \eta(1,r) &= \frac{1}{\sqrt{B(1,r)\cdot B^\dagger(1,r)}},  \quad\quad \eta(\obar,r) = \frac{1}{\sqrt{A^\dagger(\obar,r)\cdot A(\obar,r)}},
    \label{eq: eta 1 and 1 bar expressions}\\
    \theta(1,r) &= \frac{1}{\sqrt{A^\dagger(1,r)\cdot A(1,r)}},  \quad\quad \theta(\obar,r) = \frac{1}{\sqrt{B(\obar,r)\cdot B^\dagger(\obar,r)}},
    \label{eq: theta 1 and 1 bar expressions}\\
   \delta(1/\obar,r) &= \frac{1}{\sqrt{\big(A^\dagger(1/\obar,r)\cdot A(1/\obar,r) +2 \big)\;B^\dagger(1/\obar,r)\cdot B(1/\obar,r)}}.
    \label{eq: delta 1 and 1 bar expressions}
\end{align}
where $1/\obar$ indicates that the expression holds individually for both $1$ and $\obar$ direction arguments.

Putting everything together, the prepotential reformulation of the Kogut-Susskind Hamiltonian in Eq.~\eqref{eq: KS-ham} is given by
\begin{align}
    H&= H_M+H_E+ H_{I} \nonumber\\ 
    &=\mu \sum_{r=1}^{N} (-1)^r \psi^\dagger(r) \cdot \psi(r) \nonumber \\
    &\quad + \sum_{r=1}^{N'} \left( \frac{1}{3} \Big( \hat{N}_A(1,r)^2 + \hat{N}_B(1,r)^2 + \hat{N}_A(1,r) \hat{N}_B(1,r) \Big) + \hat{N}_A(1,r) + \hat{N}_B(1,r) \right) \nonumber\\
    &\quad + x\sum_{r=1}^{N'} \Bigl(
    \left[\psi^\dagger\cdot B^{\dagger}(1) \; \eta(1) \right]_r  \left[ \eta(\obar) \; \psi \cdot A^\dagger(\obar) \right]_{r+1} \nonumber  + \,\left[\psi^\dagger \cdot A(1) \; \theta(1) \right]_r  \Bigl[ \theta(\obar) \; \psi \cdot B(\obar) \Bigr]_{r+1}\nonumber\\
    &\quad \quad + \,\left[\psi^\dagger \cdot A^\dagger(1) \wedge B(1)  \; \delta(1) \right]_r \left[ \delta(\obar) \; \psi \cdot B^{\dagger}(\obar) \wedge A(\obar)\right]_{r+1} + \mathrm{H.c.}\Bigl).
    \label{eq: KS ham in prepotential final}
\end{align}
Importantly, the above Hamiltonian is written explicitly in terms of SU(3)-invariant contractions of site-local triplets.
It is also easy to see that it conserves the Abelian Gauss's laws.
In the next section, we will define a suite of SU(3)-invariant loop-string-hadron operators to ultimately derive the LSH Hamiltonian corresponding to Eq.~\eqref{eq: KS ham in prepotential final} in terms of them.
\section{SU(3) LSH framework derived from prepotentials
\label{sec: LSH-framework}
}

With all of the machinery established for both SU(3) irreducible Schwinger bosons, a staggered fermion field, and a Hamiltonian expressed in terms of manifestly SU(3)-invariant contractions at sites, the groundwork now exists to derive the loop-string-hadron formulation of the same theory. The LSH approach ultimately replaces the SU(3)-covariant fields used in the ISB formulation with fields that are intrinsically SU(3)-invariant.
This reformulation has the distinct advantage of having only local Abelian constraints to deal with (preserving the locality of the Hamiltonian that is otherwise lost when total gauge-fixing is applied), while substantially reducing the number of bosonic degrees of freedom.

Section \ref{subsubsec: LSH-operators} provides a heuristic discussion of gauge-invariant operators and excitations, which is intended to motivate the local basis that is established in Sec.~\ref{subsec: LSH-basis}.
In Sec.~\ref{subsubsec: LSH operator factorization}, with the local basis in hand, we obtain operator factorizations that make matrix elements of SU(3)-singlet operators explicit.
Sections \ref{subsubsec: LSH Ham construction} and \ref{subsubsec: LSH operators in terms of basis and AGL} provide a summary of the LSH formulation suitable for the whole lattice, including the Hamiltonian and a global-basis implementation of the operators that appear in the Hamiltonian.

\subsection{\label{subsubsec: LSH-operators}
Site-local SU(3) singlets and LSH excitations
}

An essential feature of the Hamiltonian for ISBs plus fermions is its construction in terms of site-local, SU(3)-invariant singlets, formed by suitable contractions of the ISBs and/or quark fields with the invariant tensors $\delta$ and $\epsilon$.
As we will be concentrating exclusively on operators all living at a given site $r$, from now until Sec.~\ref{subsubsec: LSH Ham construction} we shall suppress the position arguments.

Mass terms in Eq.~\eqref{eq: KS ham in prepotential final} are expressed entirely in terms of the SU(3)-invariant total fermion numbers, $ \psi^\dagger \cdot \psi$. Similarly, the Casimir operators to either side of a site given in Eq.~\eqref{eq: electric field casimir constraint}, which are necessary for the electric Hamiltonian, are expressed entirely in terms of the SU(3)-invariant Schwinger boson occupation numbers: $\hat{N}_A(1)$, $\hat{N}_B(1)$, $\hat{N}_A(\obar)$, and $\hat{N}_B(\obar)$.
Together, the five operators
\begin{align}
    \hat{N}_{\psi} \equiv \psi^\dagger \cdot \psi, \ \ \ \hat{P}(\obar) \equiv  \hat{N}_A(\obar) ,\ \ \ \hat{Q}(\obar)\equiv\hat{N}_B(\obar) , \ \ \ \hat{P}(1) \equiv \hat{N}_B(1)  , \ \ \ \hat{Q}(1) \equiv \hat{N}_A(1) , 
    \label{eq: irrep quantum numbers}
\end{align}
furnish a local complete set of commuting observables (CSCO) for the SU(3)-invariant subspace of the ISBs plus fermions. Note that the definitions of $\hat{P}$ and $\hat{Q}$ in terms of $A$- and $B$-type number operators are interchanged for the $1$ and $\obar$ sides of a site -- the reason for this choice will become apparent later.

The gauge-matter interaction, $H_I$, is distinct because it couples the fermions and gauge bosons, making it responsible for generating the Hilbert space of gauge-invariant states and for the dynamics among them.
In addition to the CSCO identified above, hopping terms in $H_I$ explicitly depend on the local SU(3) singlets
\begin{align}
    & \psi^\dagger \cdot B^\dagger(1) , \ \psi^\dagger \cdot A(1) , \ \psi^\dagger \cdot A^\dagger(1) \wedge B(1) , \  \psi^\dagger \cdot B^\dagger(\obar) , \ \psi^\dagger \cdot A(\obar) , \ \psi^\dagger \cdot A^\dagger (\obar) \wedge B(\obar) , \text{ and H.c.}
    \label{eq: Hinter singlets}
\end{align}
Local, observable configurations of quarks and gauge flux can be reached from the local fermionic and bosonic vacuum, a subspace on which all operators in the local CSCO evaluate to zero. In other words, all site-local configurations can be generated by acting on the local vacuum with the SU(3) singlets from which $H$ is constructed.
It will be shown that all local configurations can be characterized in terms of two distinct SU(3)-invariant bosonic modes and three distinct SU(3)-invariant fermionic modes. The remainder of this section is intended to heuristically motivate the associations between various SU(3)-singlet operators and distinct SU(3)-invariant fermionic and bosonic excitations, providing context for a subsequent basis ansatz that veritably covers all possible excitations at a site.

In the Kogut-Susskind formulation and the ISB formulation with fermions, there are three quark modes per site, denoted by $\psi^\dagger_\alpha$ with $\alpha=1,2,3$ for the three quark colors.
By definition, however, all of the $\psi^\dagger_\alpha$ are gauge covariant rather than gauge invariant.
Still, the total quark number $\psi^\dagger \cdot \psi$ is a gauge-invariant quantum number and can take values from 0 to 3, suggesting that there should exist three SU(3)-invariant quark modes.
The LSH approach seeks to identify and use such excitations as elementary degrees of freedom.
The first task, then, is to identify the three SU(3)-invariant fermionic excitations.

Two important singlets that are clearly needed are those in Eq.~\eqref{eq: Hinter singlets} that involve creation operators only, namely, $\psi^\dagger \cdot B^\dagger(1)$ and $\psi^\dagger \cdot B^\dagger(\obar)$, because these can create nontrivial states when applied to the local vacuum.
Taking these to correspond with distinct modes, one may associate a fermionic creation operator $\chi_1^\dagger$ ($\chi_{\obar}^\dagger$) with $\psi^\dagger \cdot B^\dagger(1)$ ($\psi^\dagger \cdot B^\dagger(\obar)$).
One then searches for a third independent mode that can be excited from the local vacuum. Such an operator must (i) be gauge invariant, (ii) contain only creation operators, and (iii) have exactly one fermionic creation operator. From Table~\ref{tab: prepotential irreps}, it can be deduced that $\psi^\dagger \cdot A^\dagger(1) \wedge A^\dagger(\obar)$ is the only such SU(3) singlet not yet enumerated.
To see that the $\psi^\dagger \cdot A^\dagger(1) \wedge A^\dagger(\obar)$ excitation is indeed required by the Hamiltonian, one first notes that
\begin{align}
    \{ \psi^\dagger \cdot A^\dagger(1) \wedge B(1) \, , \, \psi^\dagger \cdot B^\dagger(1) \} &= - \psi^\dagger \cdot \psi^\dagger \wedge A^\dagger(1) ,
\end{align}
implying the necessity of a two-quark excitation $\psi^\dagger \cdot \psi^\dagger \wedge A^\dagger(1)$, and then that
\begin{align}
    [ \psi^\dagger \cdot \psi^\dagger \wedge A^\dagger(1) \, , \, \psi \cdot A^\dagger(\obar) ] &= \psi^\dagger \cdot A^\dagger(\obar) \wedge A^\dagger(1) ,
\end{align}
where $\psi \cdot A^\dagger(\obar)$ is considered because it is one of the singlets identified in Eq.~\eqref{eq: Hinter singlets}.
Returning to the original set of singlets from Eq.~\eqref{eq: Hinter singlets}, the operators $\psi^\dagger \cdot A(1)$ and $\psi^\dagger \cdot A(\obar)$
are associated with $\hat{\chi}_{\obar}^\dagger$ and $\hat{\chi}_{1}^\dagger$, respectively, while both of $\psi^\dagger \cdot A^\dagger(1) \wedge B(1)$ and $\psi^\dagger \cdot A^\dagger(\obar) \wedge B(\obar)$ are associated with $\hat{\chi}_0^\dagger$;
these associations will be validated later, in Sec.~\ref{subsubsec: LSH operator factorization}.

In addition to the fermionic modes, there must be two bosonic excitations of the form $A^\dagger(\obar) \cdot B^\dagger(1)$ and $B^\dagger(\obar) \cdot A^\dagger(1)$ because of the relations
\begin{align}
    \{ \psi^\dagger \cdot B^\dagger(1) , \psi \cdot A^\dagger(\obar) \} &= A^\dagger(\obar) \cdot B^\dagger(1) , \\
    \{ \psi^\dagger \cdot B^\dagger(\obar) , \psi \cdot A^\dagger (1)\} &= B^\dagger(\obar) \cdot A^\dagger(1) .
\end{align}
Note that $A^\dagger(\obar) \cdot B^\dagger(1)$ [$B^\dagger (\obar)\cdot A^\dagger(1)$] increases $\hat{P}(\obar)$ and $\hat{P}(1)$ [$\hat{Q}(\obar)$ and $\hat{Q}(1)$] by one unit, and will hence be associated with a bosonic degree of freedom $\hat{n}_P$ [$\hat{n}_Q$].
Beyond these two excitations, there are no other independent, purely bosonic SU(3)-invariant excitations.
To support this claim, one may use counting arguments to deduce there are at most two unconstrained bosonic degrees of freedom:
(i) In the Kogut-Susskind formulation, the states of the adjacent link ends can each be characterized in terms of two Casimirs and three additional quantum numbers (hypercharge, isospin, and third component of isospin), giving a total of ten variables that are then subject to eight non-Abelian Gauss's law constraints.
(ii) In the prepotential plus fermions formulation, the two triplets on either side of a site add up to 12 bosonic occupation numbers, which are then subject to
eight components of Gauss's law and the two additional constraints $A^\dagger(1) \cdot B^\dagger(1) \simeq 0$ and $A^\dagger(\obar) \cdot B^\dagger(\obar) \simeq 0$.
At a more practical level, and in contrast to the counting arguments, one may alternatively note that there are only four possible bilinear contractions that can be made out of the bosonic creation operators $A^\dagger_\alpha(1)$, $A^\dagger_\alpha(\obar)$, $B^{\dagger \, \alpha}(1)$, and $B^{\dagger \, \alpha}(\obar)$.
Two of these bilinears are the bosonic excitations identified above, while the other two, $A^\dagger(1) \cdot B^\dagger(1)$ and $A^\dagger (\obar)\cdot B^\dagger(\obar)$, are equivalent to zero with the irreducible Schwinger boson construction.
One could also imagine constructing trilinears by contractions with the $\epsilon$ tensor, but any nontrivial contraction would have to involve at least one annihilation triplet.

To summarize, the operators appearing directly in the Hamiltonian can be associated with a variety of purely-creation SU(3) singlets with the following associations:
\begin{align*}
    \psi^\dagger \cdot B^\dagger(1) , \ \psi^\dagger \cdot A(\obar) \ &: \quad \text{apply } \hat{\chi}_1^\dagger , \\
    \psi^\dagger \cdot B^\dagger(\obar) , \ \psi^\dagger \cdot A(1) \ &: \quad \text{apply } \hat{\chi}_{\obar}^\dagger , \\
    \psi^\dagger \cdot A^\dagger (1)\wedge B(1) , \ \psi^\dagger \cdot A^\dagger(\obar) \wedge B(\obar) , \ \psi^\dagger \cdot A^\dagger(\obar) \wedge A^\dagger(1) \ &: \quad \text{apply } \hat{\chi}_0^\dagger , \\
    A^\dagger(\obar) \cdot B^\dagger(1) \ &: \quad \text{raise $\hat{n}_P$ by one} , \\
    B^\dagger (\obar)\cdot A^\dagger(1) \ &: \quad \text{raise $\hat{n}_Q$ by one} .
\end{align*}
In addition to these, there are also multi-quark excitations that can be regarded as composite:
\begin{align*}
    \psi^\dagger \cdot B^\dagger(\obar) \, \psi^\dagger \cdot B(1) \ &: \quad \text{apply } \hat{\chi}_{\obar}^\dagger \chi_1^\dagger \\ 
    \psi^\dagger \cdot \psi^\dagger \wedge A^\dagger(1) \ &: \quad \text{apply } \hat{\chi}_0^\dagger \chi_1^\dagger \\ 
    \psi^\dagger \cdot \psi^\dagger \wedge A^\dagger(\obar) \ &: \quad \text{apply } \hat{\chi}_{\obar}^\dagger \chi_{0}^\dagger \\
    \psi^\dagger \cdot \psi^\dagger \wedge \psi^\dagger \ &: \quad \text{apply $\hat{\chi}_{\obar}^\dagger \hat{\chi}_{0}^\dagger \chi_1^\dagger$}
\end{align*}
We note that one may take a different perspective on how to identify the set of local excitations, which proceeds by taking all available triplets at the site and listing all nontrivial bilinear and trilinear contractions of purely creation operators.
One will ultimately arrive at the same set of creation operators that cover both ``elementary'' degrees of freedom and composite excitations.

The following sections will serve to derive LSH operator factorizations for the above operators by considering a local basis and demonstrating that the factorizations behave in all ways identically to the SU(3)-invariant contractions -- at least within the space of excitations that are dynamically relevant.

\subsection{LSH basis: Single site
\label{subsec: LSH-basis}
}
\begin{figure}[t]
    \includegraphics[scale=1]{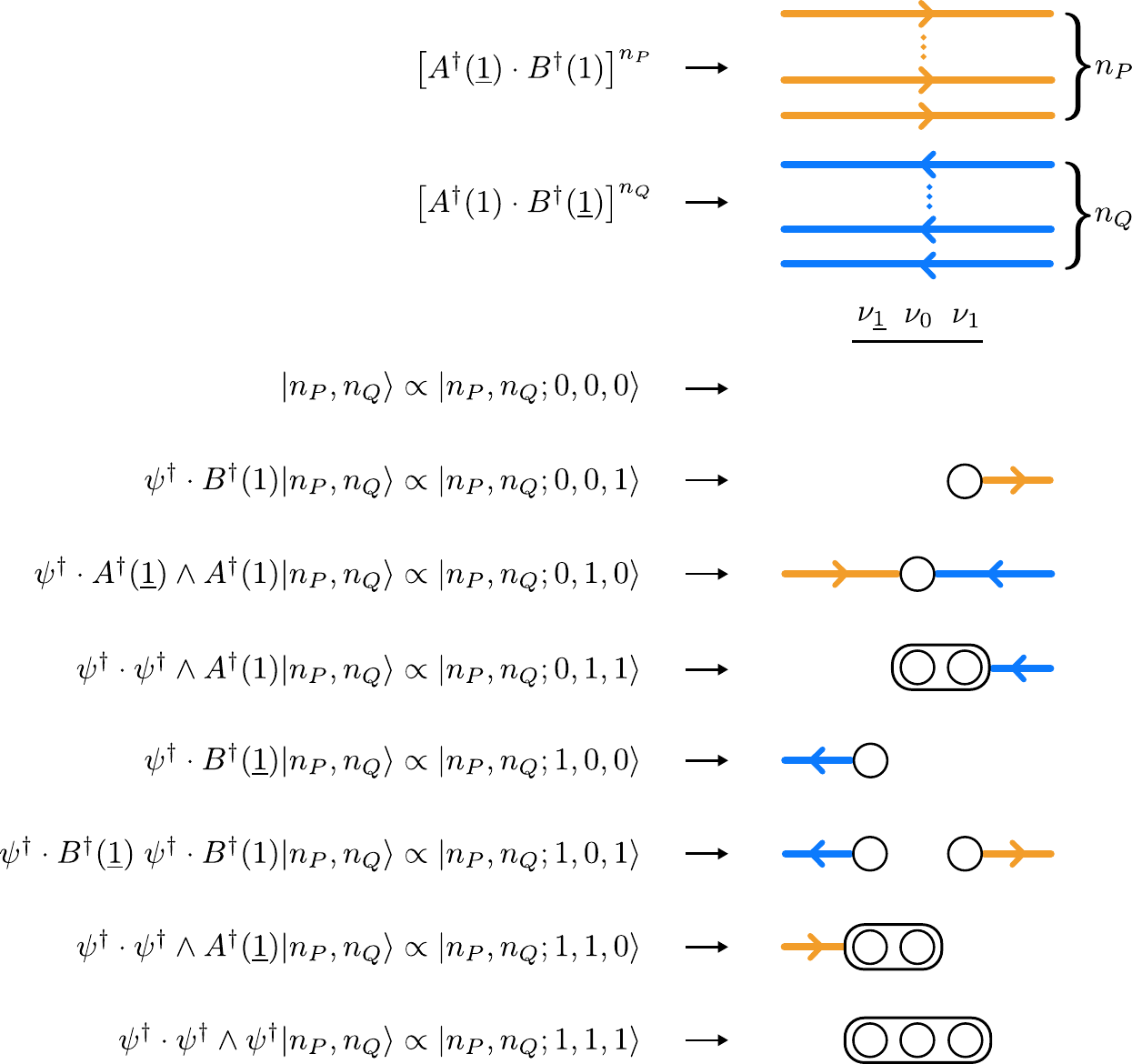}
    \caption{Pictorial representation of single-site SU(3) loop string-hadron states defined in Eqs.~\eqref{eq: LSH states unnormalized def} and \eqref{eq: Unnormalized LSH-basis-state}. At each site, electric flux of two types can pass through the site, illustrated by rightward (or orange) arrows counted by $n_P$, and leftward (blue) arrows counted by $n_Q$.
    Additionally there may exist three gauge singlet gauge-matter excitations, characterized by three fermionic occupation numbers $\nui,\num,\nuo$.
    For states with $\sum_f \nu_f = 1$, the quarks either source a unit of flux outward to one side, or they sink two incoming units of flux, one unit per side.
    For states with $\sum_f \nu_f = 2$, the quarks either source two units of outward flux, one unit per side, or they sink a single incoming unit of flux from one side.
    The $\sum_f \nu_f = 3$ configuration is the on-site baryon, which neither sources nor sinks any gauge flux.
    \label{fig: pict-LSH}}
\end{figure}
Above, the SU(3)-singlet operators relevant to expressing the Hamiltonian have been identified and arguments have been made for the existence of exactly two SU(3)-invariant bosonic excitations and three SU(3)-invariant fermionic excitations. 
In this section, we make this concrete by again restricting the discussion to one site, writing down a local-basis ansatz in terms of the two bosonic and three fermionic quantum numbers, and showing that the Hilbert space spans all possible local configurations that can be generated by Hamiltonian.
With a valid local basis identified, one can use it to convert the SU(3)-invariant contractions involving ISBs into factorized expressions that only make reference to the SU(3)-invariant degrees of freedom, following what was done for the SU(2) LSH formulation \cite{Raychowdhury:2019iki}.
To perform these calculations, one must make repeated use of 
the (anti)commutation relations obeyed by the colored fermion components and the ISBS;
the nontrivial bosonic commutation relations obeyed by ISBs, which are necessary for SU(3), make the calculations substantially more involved than the case of SU(2).
Once the factorized expressions are obtained, it is possible to express the entire formulation without any reference whatsoever to gauge-covariant quarks or ISBs.

We now propose a basis ansatz for site-local excitations that is characterized in terms of a set of gauge-invariant quantum numbers, i.e., the loop-string-hadron quantum numbers in the strong coupling basis.
The first object needed is the local normalized vacuum ket, denoted $\ket{\Omega}$, defined by the simultaneous conditions
\begin{align}
    \hat{P}(1) \ket{\Omega} = \hat{Q}(1) \ket{\Omega} = \hat{P}(\obar) \ket{\Omega} = \hat{Q}(\obar) \ket{\Omega} = \hat{N}_\psi \ket{\Omega} &= 0, \\
    \braket{\Omega | \Omega} &= 1.
\end{align}
Next, following the discussion of the preceding section, an ansatz for eight orthogonal local matter configurations $\kket{0 \, 0 \, ; \nu_{\obar} , \nu_0 , \nu_1}$ can be defined in terms of three bits $\nu_{\obar}$, $\nu_0$, $\nu_1$ as
\begin{subequations}
\begin{align}
    \kket{0 \, 0 ; 0 \, 0 \, 0} &\equiv \ket{\Omega} , \\
    \kket{0 \, 0 ; 0 \, 0 \, 1} &\equiv
    \psi^\dagger \cdot B^\dagger(1) \ket{\Omega} , \\
    \kket{0 \, 0 ; 1 \, 0 \, 0} &\equiv
    \psi^\dagger \cdot B^\dagger(\obar) \ket{\Omega} , \\
    \kket{0 \, 0 ; 0 \, 1 \, 0} &\equiv
    \psi^\dagger \cdot A^\dagger(\obar) \wedge A^\dagger(1) \ket{\Omega} , \\
    \kket{0 \, 0 ; 1 \, 0 \, 1} &\equiv
    \psi^\dagger \cdot B^\dagger(\obar) \ \psi^\dagger \cdot B^\dagger(1) \ket{\Omega} , \\
    \kket{0 \, 0 ; 0 \, 1 \, 1} &\equiv
    \tfrac{1}{2} \psi^\dagger \cdot \psi^\dagger \wedge A^\dagger(1) \ket{\Omega} , \\
    \kket{0 \, 0 ; 1 \, 1 \, 0} &\equiv 
    \tfrac{1}{2} \psi^\dagger \cdot \psi^\dagger \wedge A^\dagger(\obar) \ket{\Omega} , \\
    \kket{0 \, 0 ; 1 \, 1 \, 1} &\equiv \tfrac{1}{6} \psi^\dagger \cdot \psi^\dagger \wedge \psi^\dagger \ket{\Omega} ,
\end{align}
\label{eq: LSH states unnormalized def}
\end{subequations}
where the double-bar ket $\kket{\ }$ is used to denote that the above kets are not necessarily normalized.
The infinite towers of bosonic excitations are accommodated via 
\begin{equation}
    \kket{n_P,n_Q\,;\,\nu_{\obar},\nu_0,\nu_1} \equiv \bigl(A^\dagger(\obar)\cdot B^\dagger(1)\bigr)^{n_P} \bigl(B^\dagger(\obar)\cdot A^\dagger(1)\bigr)^{n_Q}\kket{0,0\,;\,\nu_{\obar},\nu_0,\nu_1} ,
    \label{eq: Unnormalized LSH-basis-state}
\end{equation}
where $n_P$ and $n_Q$ are nonnegative integers and the double-bar notation serves the same purpose as in Eqs.~\eqref{eq: LSH states unnormalized def}.

The above states can be represented pictorially as shown in Fig.~\ref{fig: pict-LSH}. Such a pictorial representation is often useful for obtaining an intuitive understanding of the LSH degrees of freedom, which can be described as follows:
\begin{itemize}
    \item $n_P$ and $n_Q$ are associated with two independent types of bosonic flux running through the site, without originating from or terminating on any quarks. We choose $n_P$ to count units of ``rightward'' (or orange) flux running from $\obar\to 1$, and $n_Q$ to count units of ``leftward'' (or blue) flux running from $1\to\obar$.
    \item A lone excitation of $\nu_{1}$ ($\nu_{\obar}$) sources one unit of outward flux, pointing in the $1$ (\obar) direction.
    \item The $\nu_{1}$ and $\nu_{\obar}$ excitations can be present simultaneously, with each sourcing an outward flux unit as described above.
    \item A lone excitation of $\nu_0$ sinks two units of flux, with one unit pointing inward from both sides of the site.
    \item The excitations of $\nu_0$ paired with $\nu_{1}$ ($\nu_{\obar}$) acts like an antiquark by sinking one unit of flux, pointing inward from the $1$ ($\obar$) direction.
    \item The excitation of all $\nu_{\obar}$, $\nu_0$, and $\nu_1$ modes together is an on-site baryon, which neither sources nor sinks any gauge flux (but flux units can flow through it).
\end{itemize}

Later, in Sec.~\ref{subsubsec: LSH operators in terms of basis and AGL}, we will discuss how the states in the physical Hilbert space of the whole lattice are constructed from the tensor product of local LSH states in Eq.~\eqref{eq: LSH states unnormalized def} subjected to the Abelian Gauss's law constraints (equivalent to Eq.~\eqref{eq: AGL in ISB}). These constraints are simple to visualize in terms of the pictorial representation: they translate into conservation of rightward and leftward gauge fluxes along any given link.

Finally, the above states yield an orthonormal basis given by
\begin{equation}
    \ket{n_P,n_Q\,;\,\nu_{\obar},\nu_0,\nu_1} =\mathcal N^{n_P,n_Q}_{\nu_{\obar},\nu_0,\nu_1} \bigl(A^\dagger(\obar)\cdot B^\dagger(1)\bigr)^{n_P} \bigl(B^\dagger(\obar)\cdot A^\dagger(1)\bigr)^{n_Q}\kket{0,0\,;\,\nu_{\obar},\nu_0,\nu_1} , 
    \label{eq: LSH-basis-state}
\end{equation}
where $\mathcal N^{n_P,n_Q}_{\nu_{\obar},\nu_0,\nu_1}$ are the normalization factors obtained by solving the condition
\begin{equation}
    \braket{n_P',n_Q'\,;\,\nu'_{\obar},\nu'_{0},\nu'_1|\,n_P,n_Q\,;\,\nu_{\obar},\nu_{0},\nu_1} = \delta_{n_P}^{n_P'}\; \delta_{n_Q}^{n_Q'}\; \delta_{\nu_{\obar}}^{\nu_{\obar}'}\; \delta_{\num}^{\num'}\; \delta_{\nu_1}^{\nu_1'},
    \label{eq: normalization-cond}
\end{equation}
and using the algebraic relations in Eqs.~\eqref{eq: ferm_anticomm}, and~\eqref{eq: AAdagg-commutator}-\eqref{eq: zero-commutators}.
The coefficients' expression in terms of the LSH occupation numbers turns out to be
\begin{equation}
    \mathcal {N}^{\,n_P,n_Q}_{\,\nu_{\obar},\num,\nu_1} =  \left[\frac{1}{2} (n_P+n_Q+3-\delta_{\num\,\nu_{\obar}}\delta_{\num\,\nu_{1}}) (n_P+2-\delta_{\num\,\nu_{1}})! \, (n_Q+2-\delta_{\num\,\nu_{\obar}})! \, n_P! \, n_Q! \right]^{-1/2} .
    \label{eq: normalization factor expression}
\end{equation}

With the basis ansatz above, the next step is to confirm that all possible states that are dynamically connected to the local vacuum are indeed spanned by it.
To do this, it is sufficient to confirm that the states are closed under the application of each and every SU(3)-singlet operator appearing in the Hamiltonian.
The simplest place to start is with the operators appearing in $H_M$ and $H_E$.
It is easy to see that $\hat{N}_\psi$ is diagonalized by this basis, with
\begin{align}
    \hat{N}_\psi \ket{n_P , n_Q ; \nu_{\obar}, \nu_0, \nu_1} &= (\nu_{\obar} + \nu_0 + \nu_1) \ket{n_P,n_Q\,;\,\nu_{\obar},\nu_{0},\nu_1} .
\end{align}
As for the total ISB occupation numbers to either side, $\hat{P}(1/\obar)$ and $\hat{Q}(1/\obar)$, these are also diagonalized.
Note that each factor of $A^\dagger(\obar)\cdot B^\dagger(1)$ increases $\hat{P}(1)$ and $\hat{P}(\obar)$ by one unit,
while each factor of $B^\dagger(\obar)\cdot A^\dagger(1)$ increases $\hat{Q}(1)$ and $\hat{Q}(\obar)$ by one unit.
It is for this reason that the definitions of $\hat{P}$ and $\hat{Q}$ were swapped at one end of a link relative to the other in Eq.~\eqref{eq: irrep quantum numbers} and that the symbols $n_P$ and $n_Q$ were chosen.
Careful inspection of the matter configurations reveals that the ISB occupation numbers may be increased by one additional unit, depending on the specific combination of $\nu_f$'s.
These observations are encapsulated by
\begin{subequations}
    \label{eq: P Q numbers on LSH states}
\begin{align}
    \hat{P}(1) \ket{n_P,n_Q\,;\,\nu_{\obar},\nu_{0},\nu_1} &= ( n_P +  \nu_1\left(1-\nu_0  \right) ) \ket{n_P,n_Q\,;\,\nu_{\obar},\nu_{0},\nu_1} , \\
    \hat{Q}(1) \ket{n_P,n_Q\,;\,\nu_{\obar},\nu_{0},\nu_1} &= ( n_Q +  \nu_0 (1-\nu_{\obar}  ) ) \ket{n_P,n_Q\,;\,\nu_{\obar},\nu_{0},\nu_1} , \\
    \hat{Q}(\obar) \ket{n_P,n_Q\,;\,\nu_{\obar},\nu_{0},\nu_1} &= ( n_Q +  \nu_{\obar}\left(1-\nu_0  \right) ) \ket{n_P,n_Q\,;\,\nu_{\obar},\nu_{0},\nu_1} , \\
    \hat{P}(\obar) \ket{n_P,n_Q\,;\,\nu_{\obar},\nu_{0},\nu_1} &= ( n_P +  \nu_0\left(1-\nu_1  \right) ) \ket{n_P,n_Q\,;\,\nu_{\obar},\nu_{0},\nu_1} .
\end{align}
\end{subequations}
Pictorially, the total types of each flux flowing to either side of the site are counted according to Eqs.~\eqref{eq: P Q numbers on LSH states} as follows:
\begin{itemize}
    \item $P(1)$ counts total units of flux pointing out from the $1$ side, with $n_P$ units coming from pure gauge flux, plus an additional unit if $\nu_1$ is excited, unless $\nu_0$ is also excited.
    \item $Q(\obar)$ counts total units of flux pointing out from the $\obar$ side, with $n_Q$ units coming from pure gauge flux, plus an additional unit if $\nu_{\obar}$ is excited, unless $\nu_0$ is also excited.
    \item $Q(1)$ counts total units of flux pointing in from the $1$ side, with $n_Q$ units coming from pure gauge flux, plus an additional unit if $\nu_0$ is excited, unless $\nu_{\obar}$ is also excited.
    \item $P(\obar)$ counts total units of flux pointing in from the $\obar$ side, with $n_P$ units coming from pure gauge flux, plus an additional unit if $\nu_0$ is excited, unless $\nu_{1}$ is also excited.
\end{itemize}

Turning to $H_I$, every term is strictly off-diagonal, so the real validation of the basis ansatz lies in its closure under the application of local singlets in Eq.~(\ref{eq: Hinter singlets}).
For example, in the case of $\psi^\dagger \cdot B^\dagger(1)$, one can show using Eqs.~\eqref{eq: ferm_anticomm}, \eqref{eq: AAdagg-commutator}-\eqref{eq: zero-commutators}, \eqref{eq: LSH states unnormalized def}, and \eqref{eq: normalization factor expression} that
\begin{subequations}
\begin{align}
    \label{eq: example-action-first}
    \psi^\dagger \cdot B^\dagger(1) \ket{n_P,n_Q;0 \, 0 \, 0} &= \sqrt{n_P+2} \sqrt{\frac{n_P+n_Q+3}{n_P+n_Q+2}} \ket{n_P\,n_Q;001} , \\
    \psi^\dagger \cdot B^\dagger(1) \ket{n_P,n_Q;0 \, 1 \, 0} &= - \sqrt{n_P+1} \sqrt{\frac{n_P+n_Q+4}{n_P+n_Q+3}} \ket{n_P+1,n_Q;011} , \\
    \psi^\dagger \cdot B^\dagger(1) \ket{n_P,n_Q;1 \, 0 \, 0} &= - \sqrt{n_P+2} \ket{n_P\,n_Q;101} , \\
    \psi^\dagger \cdot B^\dagger(1) \ket{n_P,n_Q;1 \, 1 \, 0} &= \sqrt{n_P+1} \ket{n_P+1,n_Q;111} , \\
    \psi^\dagger \cdot B^\dagger(1) \ket{n_P,n_Q; \nu_{\obar} \, \num \, 1} &= 0 \quad \forall\; \nu_{\obar},\,\num \, .
\end{align}
\label{eq: 1 psi-dagg B-dagg operator actions}
\end{subequations}
For its Hermitian conjugate, one has
\begin{subequations}
\begin{align}
    \psi \cdot B(1) \ket{n_P,n_Q;0\, 0\, 1} &=  \sqrt{n_P+2} \sqrt{\frac{n_P+n_Q+3}{n_P+n_Q+2}} \ket{n_P\,n_Q;000} , \\
    \psi \cdot B(1) \ket{n_P,n_Q;0\, 1\, 1} &= -\sqrt{n_P}   \sqrt{\frac{n_P+n_Q+3}{n_P+n_Q+2}} \ket{n_P-1,n_Q;010} , \\
    \psi \cdot B(1) \ket{n_P,n_Q;1\, 0\, 1} &= -\sqrt{n_P+2} \ket{n_P\,n_Q;100} , \\
    \psi \cdot B(1) \ket{n_P,n_Q;1\, 1\, 1} &=  \sqrt{n_P}   \ket{n_P-1,n_Q;110} , \\
    \psi \cdot B(1) \ket{n_P,n_Q; \nu_{\obar} \, \num \, 0} &= 0 \quad \forall\; \nu_{\obar},\,\num \, .
\end{align}
\label{eq: 2 psi B operator actions}
\end{subequations}
Hence, the basis is closed under applications of $\psi^\dagger \cdot B^\dagger(1)$ and $\psi \cdot B(1)$; symmetry arguments imply the same is true for applications of $\psi^\dagger \cdot B^\dagger(\obar)$ and $\psi \cdot B(\obar)$.
One now proceeds systematically through the remaining SU(3) singlets.
For example,
\begin{subequations}
\begin{align}
    \psi^\dagger \cdot A(1) \ket{n_P,n_Q;0 \, 0 \, 0} &= \sqrt{n_Q} \ket{n_P,n_Q-1;100} , \\
    \psi^\dagger \cdot A(1) \ket{n_P,n_Q;0 \, 0 \, 1} &= \sqrt{n_Q} \sqrt{\frac{n_P+n_Q+3}{n_P+n_Q+2}} \ket{n_P,n_Q-1;101} , \\
    \psi^\dagger \cdot A(1) \ket{n_P,n_Q;0 \, 1 \, 0} &= \sqrt{n_Q+2} \ket{n_P\,n_Q;110} , \\
    \psi^\dagger \cdot A(1) \ket{n_P,n_Q;0 \, 1 \, 1} &= \sqrt{n_Q+2} \sqrt{\frac{n_P+n_Q+3}{n_P+n_Q+2}} \ket{n_P\,n_Q;111} , \\
    \psi^\dagger \cdot A(1) \ket{n_P,n_Q;1 \, \num \, \nu_1} &= 0 \quad \forall\; \num, \, \nu_{1} ,
\end{align}
\label{eq: 3 psi-dagg A operator actions}
\end{subequations}
showing that the basis is closed under applications of $\psi^\dagger \cdot A(1)$ and, by extension, $\psi^\dagger \cdot A(\obar)$.
The same is true for their adjoints, though we omit the specific formulas.
As a final explicit example, one can apply $\psi^\dagger \cdot A^\dagger(1) \wedge B(1)$ to the basis with the following results:
\begin{subequations}
\begin{align}
    \psi^\dagger \cdot A^\dagger(1) \wedge B(1) \ket{n_P,n_Q;0 \, 0 \, 0} &= - \sqrt{n_P} \sqrt{n_Q+2} \ket{n_P-1,n_Q;010} , \\
    \psi^\dagger \cdot A^\dagger(1) \wedge B(1) \ket{n_P,n_Q;0 \, 0 \, 1} &= - \sqrt{n_P+2} \sqrt{n_Q+2} \ket{n_P\,n_Q;011} , \\
    \psi^\dagger \cdot A^\dagger(1) \wedge B(1) \ket{n_P,n_Q;1 \, 0 \, 0} &= \sqrt{n_P} \sqrt{n_Q+1} \ket{n_P-1,n_Q+1;110} , \\
    \psi^\dagger \cdot A^\dagger(1) \wedge B(1) \ket{n_P,n_Q;1 \, 0 \, 1} &= \sqrt{n_P+2} \sqrt{n_Q+1} \ket{n_P,n_Q+1;111} , \\
    \psi^\dagger \cdot A^\dagger(1) \wedge B(1) \ket{n_P,n_Q;\nu_{\obar} \, 1 \, \nu_1} &= 0 \quad \forall\; \nu_{\obar},\,\nu_{1} .
    \label{eq: example-action-last}
\end{align}
\label{eq: 4 psi-dagg A-dagg B operator actions}
\end{subequations}
The above four explicit examples, together with the analogous calculations for the operators' adjoints and exchange of $1\leftrightarrow \obar$, confirms that the basis ansatz does in fact capture all local states excited by $H_I$.

\begin{figure}[t]
    \includegraphics[scale=1]{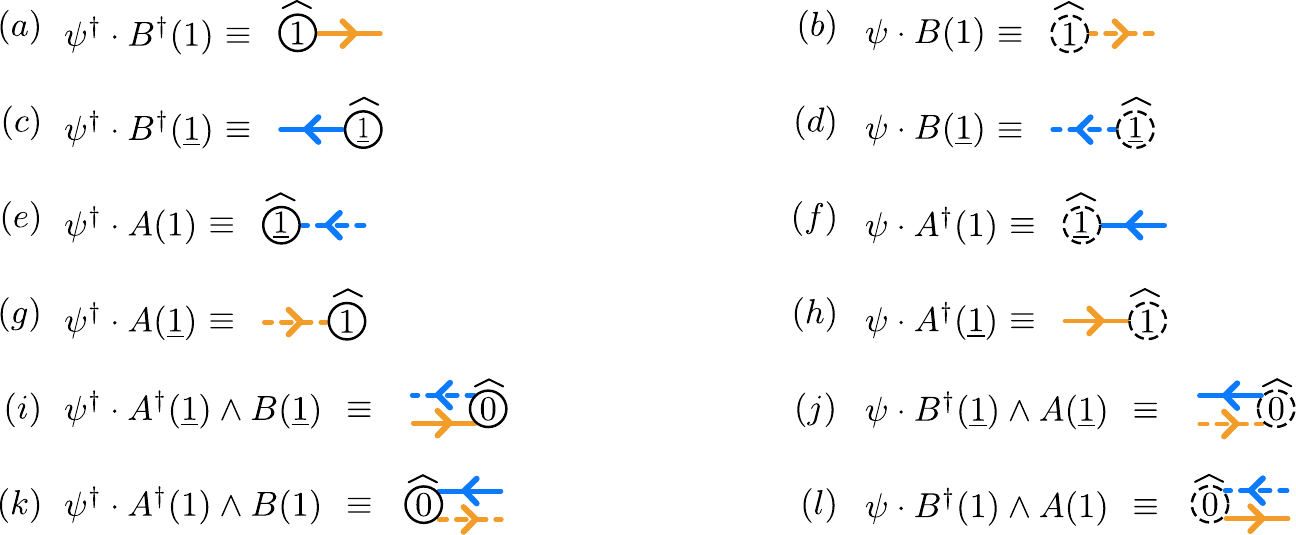}
    \caption{Pictorial representations of the LSH operators in Eq.~\eqref{eq: HI local term in prepotential} that act as local building blocks of the interaction Hamiltonian of the theory. 
    The fermion circles and arrows with lines collectively describe the creation (when solid) or annihilation (when dashed) of on-site quarks and directional gauge flux units to either side of the site.
    For a more complete explanation of this pictorial scheme, see the text in Sec.~\ref{subsec: LSH-basis}}
    \label{fig: operator actions}
\end{figure}

In Fig.~\ref{fig: pict-LSH}, a pictorial interpretation was provided for basis states of a single site.
One can also introduce a pictorial scheme for on-site SU(3) singlet operators as shown for the most important singlet operators, those appearing directly in $H_I$, in Fig.~\ref{fig: operator actions}. The rules for interpreting these pictorial representations are as following:
\begin{enumerate}
    \item A factor of ${\psi}^\dagger$ (${\psi}$) is illustrated by a solid-line (dashed-line) circle with a label $f\in \{\obar,0,1\}$ inside, corresponding to the $\nu_f$ mode it fills (empties).
    \item A factor of $A^\dagger(1)$ ($A(1)$) 
    is illustrated by a solid (dashed) arrow pointing inward from the 1 side of the site, corresponding to the creation (annihilation) of a unit of leftward gauge flux to that side.
    \item A factor of $A^\dagger(\obar)$ ($A(\obar)$) is illustrated by a solid (dashed) arrow pointing inward from the $\obar$ side of the site, corresponding to the creation (annihilation) of a unit of rightward gauge flux to that side.
    \item A factor of $B^\dagger(1)$ ($B(1)$) is illustrated by a solid (dashed) arrow pointing outward to the 1 side of the site, corresponding to the creation (annihilation) of a unit of rightward gauge flux to that side.
    \item A factor of $B^\dagger(\obar)$ ($B(\obar)$) is illustrated by a solid (dashed) arrow pointing outward to the $\obar$ side of the site, corresponding to the creation (annihilation) of a unit of leftward gauge flux to that side.
    \item A hat may be added to the drawing to distinguish it as an operator rather than a state configuration.
\end{enumerate}
Later, when multiple lattice sites are discussed, operator cartoons from different sites may be stitched together to represent more complex SU(3)-invariant operators that also manifestly preserve (or break) the Abelian Gauss's laws along links.

\subsection{\label{sec: opfac}
LSH operator factorizations: Single site
\label{subsubsec: LSH operator factorization}}
Above, a local on-site basis characterized by strictly SU(3)-invariant quantum numbers has been given and we have argued that they span the space relevant to gauge-invariant time evolution.
In the process, a number of matrix elements for the various SU(3)-singlet contractions are computed by applying the singlets of Eq.~(\ref{eq: Hinter singlets}) to basis kets and using the algebra of SU(3) ISBs and the colored fermion components.
To synthesize these results and depart from their expression in terms of SU(3) ISBs, it is now appropriate to introduce, following the original SU(2) LSH construction \cite{Raychowdhury:2019iki}, the normalized ladder operators and diagonal functions that are sufficient to express LSH dynamics without any reference to underlying degrees of freedom that transform under SU(3).

The first and most obvious objects are number operators for the LSH occupation numbers, $\hat{n}_l$ and $\hat{\nu}_f$ for the bosonic modes $l\in \{P,Q\}$ and the fermionic modes $f\in\{\obar,0,1\}$. Being diagonal in the LSH occupation number basis, these satisfy
\begin{align}
    \hat{n}_l \ket{n_P , n_Q ; \nu_{\obar}, \nu_0, \nu_1} &= n_l \ket{n_P , n_Q ; \nu_{\obar}, \nu_0, \nu_1}, \\
    \hat{\nu}_f \ket{n_P , n_Q ; \nu_{\obar}, \nu_0, \nu_1} &= \nu_f \ket{n_P , n_Q ; \nu_{\obar}, \nu_0, \nu_1}.
\end{align}
Next are the SU(3)-invariant quark modes, $\hat{\chi}_f$.
These are defined as canonical fermionic operators satisfying
\begin{align}
    \bigl\{ \hat{\chi}_{f}, \hat{\chi}_{f'} \bigr\} = \bigl\{ \hat{\chi}_{f}^{\dagger}, \hat{\chi}_{f'}^{\dagger} \bigr\} &= 0\,, \\
    \bigl\{ \hat{\chi}_{f}, \hat{\chi}_{f'}^{\dagger} \bigr\} &= \delta_{f \, f'}\,, \\
    \hat{\chi}_f \ket{n_P , n_Q \,;\, 0, 0, 0} &= 0\,.
\end{align}
They also raise or lower the $\hat{\nu}_f$ quantum numbers, as expressed by
\begin{align}
    [\hat{\nu}_{f}, \hat{\chi}_{f'}^\dagger] &= \delta_{f \, f'} \hat{\chi}_{f}^\dagger\,, \\
    [\hat{\nu}_{f}, \hat{\chi}_{f'}] &= - \delta_{f \, f'} \hat{\chi}_{f}\,.
\end{align}
Equivalently, $\hat{\nu}_f = \hat{\chi}_f^\dagger \hat{\chi}_f$.

Then there are normalized bosonic ladder operators $\hat{\ladder}_l$,
where $\hat{\ladder}_l$ ($\hat{\ladder}_l^\dagger$) lowers (raises) the bosonic quantum number $n_l$ by one unit, without rescaling a state's coefficients, aside from $\hat{\ladder}_l$ destroying the basis kets with $n_l=0$.
Algebraically,
\begin{align}
    [\hat{n}_{l} , \hat{\ladder}_{l'}^\dagger ] &= \delta_{l \, l'} \hat{\ladder}_{l}^\dagger\,, \\
    [\hat{n}_{l} , \hat{\ladder}_{l'} ] &= - \delta_{l \, l'} \hat{\ladder}_{l} \,,\\
    [\hat{\ladder}_{l} , \hat{\ladder}_{l'}^\dagger ] &= \delta_{l \, l'} \int_{-\pi}^{\pi}  \frac{d\phi}{2\pi} e^{i \phi \hat{n}_{l} } \,.
\end{align}
Above, $\int_{-\pi}^{\pi}  \frac{d\phi}{2\pi} e^{i \phi \hat{n}_{l}}$ is used to represent a projection operator onto the local bosonic $l$-mode vacuum. The bosonic ladder operators also do not ``talk'' to the fermionic modes whatsoever:
\begin{align}
    [ \hat{\nu}_{f} , \hat{\ladder}_{l} ] = [ \hat{\chi}_{f} , \hat{\ladder}_{l} ] = [ \hat{\chi}_{f}^\dagger , \hat{\ladder}_{l} ] &= 0 \,.
\end{align}
Equipped with the normalized fermionic and bosonic ladder operators, one can construct the local orthonormal basis by applying the raising operators on the bosonic and fermionic vacuum as
\begin{align}
    \ket{n_P, n_Q\,;\,\nu_{\obar},{\nu_0},\nu_1} &\quad\mapsto\quad ( \hat{\ladder}_{P}^{\dagger} ) ^{n_P}
    ( \hat{\ladder}_{Q}^{\dagger} ) ^{n_Q}
    ( \hat{\chi   }_{\obar}^{\dagger} ) ^{\nu_{\obar}}
    ( \hat{\chi   }_{0}^{\dagger} ) ^{\nu_0}
    ( \hat{\chi   }_{1}^{\dagger} ) ^{\nu_1} \ket{0,0\,;\,0,0,0} .
    \label{eq: basis in term of creation operators}
\end{align}
This local-basis definition involves choosing an order for the fermionic creation operators, although the final results of this section (the operator factorizations) are independent of this choice.

The last structures that are useful to define are those of diagonal functions and conditional (bosonic) ladder operators.
Diagonal functions are any operators that can be written in closed form in terms of only LSH number operators.
For example, the diagonal function $\sqrt{\hat{n}_l + \hat{\nu}_f}$ is defined by the matrix elements
\begin{align*}
    \bra{n_P',n_Q'\,;\,\nu'_{\obar},\nu'_{0},\nu'_1} \sqrt{\hat{n}_l + \hat{\nu}_f} \ket{n_P , n_Q ; \nu_{\obar}, \nu_0, \nu_1} &= \sqrt{n_l + \nu_f} \ \delta_{n_P}^{n_P'} \, \delta_{n_Q}^{n_Q'} \, \delta_{\nu_{\obar}}^{\nu_{\obar}'} \, \delta_{\num}^{\num'} \, \delta_{\nu_1}^{\nu_1'}  \,.
\end{align*}
The conditional ladder operators are essentially controlled operations.
For example, the expression $( \hat{\ladder}_l )^{  \hat{\nu}_f} $ is used for an operator that, when applied to kets, lowers $n_l$ if $\nu_f=1$ and does nothing if $\nu_f=0$.
Formulas for the various conditional ladder operators that arise are as follows:
\begin{alignat}{4}
    ( \hat{\ladder}_l )^{  \hat{\nu}_f} &\equiv \hat{\ladder}_l \, \hat{\nu}_f + ( 1 - \hat{\nu}_f ), \qquad& ( \hat{\ladder}_l^{\dagger} )^{  \hat{\nu}_f} &\equiv \hat{\ladder}_l^{\dagger} \, \hat{\nu}_f + ( 1 - \hat{\nu}_f ), \\
    ( \hat{\ladder}_l )^{ 1 - \hat{\nu}_f} &\equiv \hat{\ladder}_l \, ( 1 - \hat{\nu}_f )  + \hat{\nu}_f, & ( \hat{\ladder}_l^{\dagger} )^{ 1 - \hat{\nu}_f} &\equiv \hat{\ladder}_l^{\dagger} \, ( 1 - \hat{\nu}_f ) + \hat{\nu}_f .
\end{alignat}

All of the necessary ingredients to express the LSH formulation are now available to give operator factorizations for the SU(3) singlets of Eq.~(\ref{eq: Hinter singlets}) as follows:
\begin{subequations}
\label{eq: Hinter opfacs}
\begin{align}
    \psi^\dagger \cdot B^\dagger(1) &\quad\mapsto\quad \hat{\chi}_1^\dagger ( \hat{\ladder}_P^\dagger )^{\hnum}
    \sqrt{\hat{n}_P+2 - \hnum} \sqrt{\frac{\hat{n}_P+\hat{n}_Q+3+\hnum}{\hat{n}_P+\hat{n}_Q+2+\hnui+\hnum}} \\
    \psi \cdot B(1) &\quad\mapsto\quad \hat{\chi}_1 ( \hat{\ladder}_P )^{\hnum}
    \sqrt{\hat{n}_P+2 (1- \hnum)} \sqrt{\frac{\hat{n}_P+\hat{n}_Q+3}{\hat{n}_P+\hat{n}_Q+2+\hnui}} \\
    \psi^\dagger \cdot B^\dagger(\obar) &\quad\mapsto\quad \hat{\chi}_{\obar}^\dagger ( \hat{\ladder}_Q^\dagger )^{\hnum}
    \sqrt{\hat{n}_Q+2 - \hnum} \sqrt{\frac{\hat{n}_P+\hat{n}_Q+3+\hnum}{\hat{n}_P+\hat{n}_Q+2+\hat{\nu}_1+\hnum}} \\
    \psi \cdot B(\obar) &\quad\mapsto\quad \hat{\chi}_{\obar} ( \hat{\ladder}_Q )^{\hnum}
    \sqrt{\hat{n}_Q+2 (1- \hnum)} \sqrt{\frac{\hat{n}_P+\hat{n}_Q+3}{\hat{n}_P+\hat{n}_Q+2+\hat{\nu}_1}} \\
    \psi^\dagger \cdot A(1)         &\quad\mapsto\quad \hat{\chi}_{\obar}^\dagger ( \hat{\ladder}_Q )^{1-\hnum}
    \sqrt{\hat{n}_Q+2 \,\hnum} \sqrt{\frac{\hat{n}_P+\hat{n}_Q+2+\hat{\nu}_1}{\hat{n}_P+\hat{n}_Q+2}} \\
    \psi \cdot A^\dagger(1) &\quad\mapsto\quad \hat{\chi}_{\obar} ( \hat{\ladder}_Q^\dagger )^{1-\hnum}
    \sqrt{\hat{n}_Q+1+\hnum} \sqrt{\frac{\hat{n}_P+\hat{n}_Q+3-\hnum+\hat{\nu}_1}{\hat{n}_P+\hat{n}_Q+3-\hnum}} \\
    \psi^\dagger \cdot A(\obar)         &\quad\mapsto\quad \hat{\chi}_1^\dagger ( \hat{\ladder}_P )^{1-\hnum}
    \sqrt{\hat{n}_P+2 \,\hnum} \sqrt{\frac{\hat{n}_P+\hat{n}_Q+2+\hnui}{\hat{n}_P+\hat{n}_Q+2}} \\
    \psi \cdot A^\dagger(\obar) &\quad\mapsto\quad \hat{\chi}_1 ( \hat{\ladder}_P^\dagger )^{1-\hnum}
    \sqrt{\hat{n}_P+1+\hnum} \sqrt{\frac{\hat{n}_P+\hat{n}_Q+3-\hnum+\hnui}{\hat{n}_P+\hat{n}_Q+3-\hnum}} \\
    \psi^\dagger \cdot A^\dagger(1) \wedge B(1) &\quad\mapsto\quad - \hat{\chi}_0^\dagger ( \hat{\ladder}_P )^{1-\hat{\nu}_1} ( \hat{\ladder}_Q^\dagger )^{\hnui}
    \sqrt{\hat{n}_P+2 \,\hat{\nu}_1} \sqrt{\hat{n}_Q+2 - \hnui} \\
    \psi \cdot B^\dagger(1) \wedge A(1) &\quad\mapsto\quad \hat{\chi}_0 ( \hat{\ladder}_P^\dagger )^{1-\hat{\nu}_1} ( \hat{\ladder}_Q )^{\hnui}
    \sqrt{\hat{n}_P+1+\hat{\nu}_1} \sqrt{\hat{n}_Q+2(1-\hnui)} \\
    \psi^\dagger \cdot A^\dagger(\obar) \wedge B(\obar) &\quad\mapsto\quad \hat{\chi}_0^\dagger ( \hat{\ladder}_P^\dagger )^{\hat{\nu}_1} ( \hat{\ladder}_Q )^{1-\hnui}
    \sqrt{\hat{n}_P+2 - \hat{\nu}_1} \sqrt{\hat{n}_Q+2 \,\hnui} \\
     \psi \cdot B^\dagger(\obar) \wedge A(\obar) &\quad\mapsto\quad - \hat{\chi}_0 ( \hat{\ladder}_P )^{\hat{\nu}_1} ( \hat{\ladder}_Q^\dagger )^{1-\hnui}
    \sqrt{\hat{n}_P+2(1-\hat{\nu}_1)} \sqrt{\hat{n}_Q+1+\hnui}
\end{align}
\label{eq: on site operator factorinzation}
\end{subequations}
To obtain the set of equations in Eq.~\eqref{eq: on site operator factorinzation}, we performed a similar analysis as Eqs.~\eqref{eq: 1 psi-dagg B-dagg operator actions}-\eqref{eq: 4 psi-dagg A-dagg B operator actions} for all operators in Eq.~\eqref{eq: Hinter singlets}, and assumed the basis could be written in the form of Eq.~\eqref{eq: basis in term of creation operators}. Note that a different fermionic ordering in Eq.\eqref{eq: basis in term of creation operators} would lead to equivalent operator factorizations as in Eq.~\eqref{eq: Hinter opfacs} but they must satisfy all the same algebraic relations.
We stress that it is the algebraic relations of the operators and the characterization of the local vacuum that are the defining features of the LSH formulation.
For completeness, analogous factorizations for the other SU(3) singlets used to construct the local Hilbert space as in Eqs.~\eqref{eq: LSH states unnormalized def}-\eqref{eq: Unnormalized LSH-basis-state} can be found in Appendix \ref{app: extra factorizations}.

\subsection{
Complete LSH formulation including Hamiltonian and Abelian symmetries
\label{subsubsec: LSH Ham construction}}
The construction of the LSH operators and the LSH basis has thus far been focused on a single site. To complete the formulation, the framework will be adapted to the whole lattice and the Hamiltonian will be expressed.

The bosonic occupation numbers $\hat{n}_l$, bosonic lowering operators $\hat{\ladder}_l$, fermionic occupation numbers $\hat{\nu}_f$, and fermionic annihilation operators $\hat{\chi}_f$ (along with their conjugates) are elevated to site-dependent fields:
\begin{align}
\hat{n}_l \to \hat{n}_l(r), \quad  \hat{\nu}_f \to \hat{\nu}_f(r), \quad \hat{\ladder}_l \to \hat{\ladder}_l(r), \quad \hat{\chi}_f \to \hat{\chi}_f(r), \qquad (r=1,2,\cdots,N) .
\end{align}
Operators belonging to the same site obey the commutation and anticommutation relations as they were given for a single site in Sec.~\ref{subsubsec: LSH operator factorization}.
Operators belonging to different sites, however, have vanishing commutators or anticommutators as appropriate for their statistics. The algebra of LSH occupation numbers and ladder operators across the entire lattice is summarized by
\begin{align}
    [\hat{\nu}_{f} (r), \hat{\nu}_{f'} (r')] = [\hat{n}_{l} (r), \hat{n}_{l'} (r')] = [\hat{\nu}_{f} (r), \hat{n}_{l} (r')] &= 0 , \\
    [\hat{\chi}_{f} (r) , \hat{n}_{l} (r')] = [\hat{\nu}_{f} (r), \hat{\ladder}_{l} (r')] = [\hat{\chi}_{f} (r), \hat{\ladder}_{l} (r')] = [\hat{\chi}_{f}^\dagger (r), \hat{\ladder}_{l} (r')] &= 0 , \\
    \bigl\{ \hat{\chi}_{f} (r), \hat{\chi}_{f'} (r') \bigr\} &= 0 , \\
    \bigl\{ \hat{\chi}_{f} (r), \hat{\chi}_{f'}^{\dagger} (r') \bigr\} &= \delta_{f \, f'} \delta_{r \, r'} , \\
    [\hat{\nu}_{f} (r), \hat{\chi}_{f'} (r')] &= - \delta_{f \, f'} \delta_{r \, r'} \hat{\chi}_{f} (r) , \\
    [\hat{n}_{l} (r), \hat{\ladder}_{l'} (r')] &= - \delta_{l \, l'} \delta_{r \, r'} \hat{\ladder}_{l} (r) , \\
    [\hat{\ladder}_{l} (r), \hat{\ladder}_{l'}^\dagger (r')] &= \delta_{l \, l'} \delta_{r \, r'} \int_{-\pi}^{\pi}  \frac{d\phi}{2\pi} e^{i \phi \hat{n}_{l}(r) } ,
\end{align}
where $r$ and $r'$ take values from 1 to $N$.
Above, $\int_{-\pi}^{\pi}  \frac{d\phi}{2\pi} e^{i \phi \hat{n}_{l}(r)}$ is used to represent a projection operator onto the local bosonic vacuum at site $r$, which is a subspace of the lattice's Hilbert space.
Next, the fermionic modes, normalized ladder operators, and diagonal functions may be put together to express the Hamiltonian.
One should first note that the SU(3)-invariant quantum numbers of the underlying prepotentials plus staggered fermions formulation (Eq.~\ref{eq: irrep quantum numbers}) are translated as
\begin{align}
    \hat{N}_\psi (r) \mapsto &\ \hnui(r) + \hnum(r) + \hnuo(r) , \\
    \label{eq: P Q 1bar side in LSH variables}
    \hat{P}(\obar,r) \mapsto \hat{n}_P(r) + \hnum(r) (1 - \hnuo(r)), &\quad \hat{Q}(\obar,r) \mapsto \hat{n}_Q(r) + \hnui(r) (1 - \hnum(r)) , \\
    \label{eq: P Q 1 side in LSH variables}
    \hat{P}(1,r) \mapsto \hat{n}_P(r) + \hnuo(r) (1 - \hnum(r)), &\quad \hat{Q}(1,r) \mapsto \hat{n}_Q(r) + \hnum(r) (1 - \hnui(r)) .
\end{align}
Similarly, in light of Eqs.~\eqref{eq: on site operator factorinzation} the operators in Eq.~\eqref{eq: Hinter singlets} for a lattice site $r$ have the following operator factorizations:
\begin{subequations}
\begin{align}
    \psi^\dagger(r) \cdot B^\dagger(1,r) &= \left[\hat{\chi}_1^\dagger ( \hat{\ladder}_P^\dagger )^{\hnum}
    \sqrt{\hat{n}_P+2 - \hnum} \sqrt{\frac{\hat{n}_P+\hat{n}_Q+3+\hnum}{\hat{n}_P+\hat{n}_Q+2+\hnui+\hnum}}\ \right]_r \,,\\
    \psi(r) \cdot B(1,r) &= \left[\hat{\chi}_1 ( \hat{\ladder}_P )^{\hnum}
    \sqrt{\hat{n}_P+2 (1- \hnum)} \sqrt{\frac{\hat{n}_P+\hat{n}_Q+3}{\hat{n}_P+\hat{n}_Q+2+\hnui}}\ \right]_r \,, \\
    \psi^\dagger(r) \cdot A(1,r) &= \left[\hat{\chi}_{\obar}^\dagger ( \hat{\ladder}_Q )^{1-\hnum}
    \sqrt{\hat{n}_Q+2 \,\hnum} \sqrt{\frac{\hat{n}_P+\hat{n}_Q+2+\hat{\nu}_1}{\hat{n}_P+\hat{n}_Q+2}}\ \right]_r \,,\\
    \psi(r) \cdot A^\dagger(1,r) &= \left[\hat{\chi}_{\obar} ( \hat{\ladder}_Q^\dagger )^{1-\hnum}
    \sqrt{\hat{n}_Q+1+\hnum} \sqrt{\frac{\hat{n}_P+\hat{n}_Q+3-\hnum+\hat{\nu}_1}{\hat{n}_P+\hat{n}_Q+3-\hnum}}\ \right]_r \,,\\
    \psi^\dagger(r) \cdot A^\dagger(1,r) \wedge B(1,r) &= -\left[ \hat{\chi}_0^\dagger ( \hat{\ladder}_P )^{1-\hat{\nu}_1} ( \hat{\ladder}_Q^\dagger )^{\hnui}
    \sqrt{\hat{n}_P+2 \,\hat{\nu}_1} \sqrt{\hat{n}_Q+2 - \hnui} \right]_r\\
    \psi(r) \cdot B^\dagger(1,r) \wedge A(1,r) &= \left[\hat{\chi}_0 ( \hat{\ladder}_P^\dagger )^{1-\hat{\nu}_1} ( \hat{\ladder}_Q )^{\hnui}
    \sqrt{\hat{n}_P+1+\hat{\nu}_1} \sqrt{\hat{n}_Q+2(1-\hnui)} \right]_r
\end{align}
\end{subequations}
The factorizations for $\psi^\dagger(r) \cdot B^\dagger(\obar,r)$ and $\psi^\dagger(r) \cdot A(\obar,r)$ are obtained from their 1-side counterparts by interchanging the labels $P\leftrightarrow Q$ and $1\leftrightarrow \obar$.
The same is true for obtaining $\psi^\dagger(r) \cdot A^\dagger(\obar,r) \wedge B(\obar,r)$ and $\psi(r) \cdot B^\dagger(\obar,r) \wedge A(\obar,r)$, except one must also apply an overall minus sign in both cases.

One notes that the hopping Hamiltonian of Eq.~(\ref{eq: KS ham in prepotential final}) also involves the $\eta(1/\obar)$, $\theta(1/\obar)$, and $\delta(1/\obar)$ operators, which are diagonal functions in the sense defined above.
These can be translated from Eqs.~\eqref{eq: eta 1 and 1 bar expressions}-\eqref{eq: delta 1 and 1 bar expressions} into diagonal functions of $\hat{P}$ and $\hat{Q}$ operators via Eq.~\eqref{eq: irrep quantum numbers}, resulting in
\begin{alignat}{4}
    \eta(1,r) &= \frac{1}{\sqrt{\hat{P}(1,r)+2}}\sqrt{\frac{\hat{P}(1,r)+\hat{Q}(1,r)+3}{\hat{P}(1,r)+\hat{Q}(1,r)+2}}, & \hspace{18pt} \eta(\obar,r) &= \frac{1}{\sqrt{\hat{P}(\obar,r)}}, \\
    \theta(\obar,r) &= \frac{1}{\sqrt{\hat{Q}(\obar,r)+2}}\sqrt{\frac{\hat{P}(\obar,r)+\hat{Q}(\obar,r)+3}{\hat{P}(\obar,r)+\hat{Q}(\obar,r)+2}}, & \theta(1,r) &= \frac{1}{\sqrt{\hat{Q}(1,r)}}, \\
    \delta(1,r) &= \frac{1}{\sqrt{(\hat{Q}(1,r)+2)\hat{P}(1,r)}}, & \delta(\obar,r), &= \frac{1}{\sqrt{(\hat{P}(\obar,r)+2)\hat{Q}(\obar,r)}} .
\end{alignat}
Note that the $\hat{P}$ or $\hat{Q}$ operators technically have zero in their spectrum,
making their inverse square roots formally undefined.
However, they cannot actually encounter singularities because the link operator in Eq.~\eqref{eq: link operator in prepotential} that defines the $\eta(1/\obar)$, $\theta(1/\obar)$, and $\delta(1/\obar)$ operators is non-singular.

Once the $\eta$, $\theta$, and $\delta$ operators are multiplied with the SU(3) singlets of Eq.~(\ref{eq: Hinter singlets}) as dictated by Eq.~(\ref{eq: KS ham in prepotential final}), the full LSH Hamiltonian can be expressed as $H=H_M+H_E+H_I$, with
\begin{align}
    H_M = \sum_{r=1}^{N} H_M(r) &\equiv \mu \sum_{r=1}^{N} (-1)^r (\hnui(r) + \hnum(r) + \hnuo(r)) , 
    \label{eq: HM in LSH operators}\\
    H_E = \sum_{r=1}^{N'} H_E(r) &\equiv \sum_{r=1}^{N'} \frac{1}{3} \left( \hat{P}(1,r)^2 + \hat{Q}(1,r)^2 + \hat{P}(1,r) \hat{Q}(1,r) \right) + \hat{P}(1,r) + \hat{Q}(1,r) , 
    \label{eq: HE in LSH operators}\\
    H_I = \sum_{r=1}^{N'} H_I(r) &\equiv \sum_{r} x \left[ 
    \hat{\chi}_{1}^\dagger ( \hat{\ladder}_P^\dagger )^{\hnum}
    \sqrt{ 1  - \hnum/(\hat{n}_P + 2)} \sqrt{ 1 - \hnui/(\hat{n}_P+\hat{n}_Q+3)}
    \ \right]_{r} \nonumber \\
    & \hspace{1.2cm} \otimes
    \left[
    \sqrt{ 1  + \hnum/(\hat{n}_P + 1)} \sqrt{ 1 + \hnui/(\hat{n}_P+\hat{n}_Q+2)} \,
    \hat{\chi}_{1} ( \hat{\ladder}_P^\dagger )^{1-\hnum}
    \right]_{r+1} \nonumber \\
    &\quad + x \left[
    \hat{\chi}_{\obar}^\dagger ( \hat{\ladder}_Q )^{1-\hnum}
    \sqrt{ 1 + \hnum/(\hat{n}_Q + 1)}\sqrt{ 1 + \hat{\nu}_{1}/(\hat{n}_P+\hat{n}_Q+2)}
    \ \right]_{r} \nonumber \\
    &\hspace{1.2cm} \otimes \left[
    \sqrt{ 1 - \hnum/(\hat{n}_Q + 2)} \sqrt{ 1 - \hat{\nu}_{1}/(\hat{n}_P+\hat{n}_Q+3)} \, \hat{\chi}_{\obar} ( \hat{\ladder}_Q )^{\hnum}
    \right]_{r+1} \nonumber \\
    &\quad + x \left[
    \hat{\chi}_0^\dagger ( \hat{\ladder}_P )^{1-\hat{\nu}_{1}} ( \hat{\ladder}_Q^\dagger )^{\hnui}
    \sqrt{ 1 + \hat{\nu}_{1}/({\hat{n}_P+1}}) \sqrt{ 1 - {\hnui}/({\hat{n}_Q+2})}
    \ \right]_{r} \nonumber \\
    & \hspace{1.2cm} \otimes \left[
    \sqrt{ 1 - {\hat{\nu}_{1}}/({\hat{n}_P+2}}) \sqrt{ 1 + {\hnui}/({\hat{n}_Q+1}}) \,
    \hat{\chi}_0 ( \hat{\ladder}_P )^{\hat{\nu}_{1}} ( \hat{\ladder}_Q^\dagger )^{1-\hnui}
    \right]_{r+1} + \mathrm{H.c.} 
    \label{eq: HI in LSH operators}
\end{align}
This is the final Hamiltonian operator for the LSH formulation of (1+1)-dimensional SU(3) lattice gauge theory with staggered fermions, as reported in Sec.~\ref{sec: results}.

The LSH Hamiltonian, while having SU(3) symmetry intrinsically built into every constituent operator, features several remnant Abelian symmetries.
First there are three global U(1) symmetries generated by $\sum_r \hat{\nu}_f(r)$ for each type $f$, implying conservation of total fermions of each type.
These quantum numbers were discussed in more detail in Sec.~\ref{subsec: super-selection}.
Less obviously, the Hamiltonian commutes with all required Abelian Gauss's law constraints:
\begin{align}
    \label{eq: LSH Hamiltonian commutes with AGL}
    [ H \, , \, \hat{P}(1,r)-\hat{P}(\obar,r+1) ] &= [ H \, , \, \hat{Q}(1,r)-\hat{Q}(\obar,r+1) ] =  0
\end{align}
for each link $r$, where $\hat{P}$ and $\hat{Q}$ operators are as defined in Eqs.~\eqref{eq: P Q 1bar side in LSH variables}-\eqref{eq: P Q 1 side in LSH variables}.
Physical states are restricted to the simultaneous kernel of all Abelian Gauss's law generators.

\subsection{
Explicit basis implementation of the LSH Hamiltonian \label{subsubsec: LSH operators in terms of basis and AGL}
}
For classical computation or digital quantum computation, it is helpful to have an explicit basis representation of the Hilbert space and Hamiltonian.
The Hilbert space can be constructed as a tensor product space of local Hilbert spaces all in an occupation number basis:
\begin{align}
    \label{eq: LSH global basis ket}
    \otimes_{r} \ket{n_P(r), n_Q(r) \,;\; \nui(r), \num(r) , \nuo(r)} &\equiv \otimes_{r} \ket{n_P, n_Q \,;\; \nui, \num , \nuo}_r
\end{align}
where, for brevity, the $r$-dependences of the $n_l$ and $\nu_f$ quantum numbers have been collected into the single subscript $r$ on the ket.
The above states constitute what can be referred to as the ``full'' Hilbert space in the LSH formulation, i.e., the Hilbert space that contains physical and unphysical states.
The subspace relevant to physical dynamics, per Eq.~\eqref{eq: LSH Hamiltonian commutes with AGL}, is restricted to those satisfying
\begin{align}
    n_P(r) +  \nu_1(r)\left(1-\nu_0 (r) \right) &= n_P(r+1) +  \nu_0(r+1) \left(1-\nu_1 (r+1) \right) , \\
    n_Q(r) +  \nu_0(r)\left(1-\nu_{\obar} (r) \right) &= n_Q(r+1) +  \nu_{\obar}(r+1)\left(1-\nu_0 (r+1) \right) .
\end{align}
This constraint is a realization of the Abelian Gauss's law constraints originally given for ISBs Eq.~\eqref{eq: AGL in ISB}.
The Abelian Gauss's laws may be stated as: the total number of flux units flowing in either direction (rightward or leftward) is preserved from one side of a link to the other.
This was illustrated in Fig.~\ref{fig: AGL_pict}.

In practice, for numerical calculations a truncation may be necessary for finite dimensionality of the Hilbert space, for which one may introduce a cutoff $\Lambda$ on all bosonic $n_l$ quantum numbers.
In the case of open boundary conditions, $P(\obar,1)$ and $Q(\obar,1)$ are static, and the Abelian Gauss's laws constrain $P(1,N)\leq P(\obar,1) + N$ and $Q(1,N)\leq Q(\obar,1) + N$ such that it is possible to truncate the basis without necessarily truncating the dynamically relevant Hilbert space.

In the occupation number basis above, number operators are explicitly given by
\begin{subequations}
\begin{align}
    \hat{n}_l (r) &= \sum_{n_P, n_Q,\nu_{\obar} , \nu_0 , \nu_1} n_l(r) \ket{n_P,n_Q\,;\,\nu_{\obar},\nu_0,\nu_1}\bra{n_P,n_Q\,;\,\nu_{\obar},\nu_0,\nu_1}_r, \\
    \hat{\nu}_f (r) &= \sum_{n_P, n_Q,\nu_{\obar} , \nu_0 , \nu_1} \nu_{f}(r) \ket{n_P,n_Q\,;\,\nu_{\obar},\nu_0,\nu_1} \bra{n_P,n_Q\,;\,\nu_{\obar},\nu_0,\nu_1}_r,
\end{align}
\end{subequations}
The bosonic lowering operators are defined by
\begin{subequations}
\begin{align}
    \hat{\ladder}^{}_{P} (r) &= \sum_{n_P=1}^\infty \ \sum_{n_Q,\nu_{\obar} , \nu_0 , \nu_1} \ket{n_P-1, n_Q\,;\,\nu_{\obar},\nu_0,\nu_1} \bra{n_P,n_Q\,;\,\nu_{\obar},\nu_0,\nu_1}_r, \\
    \hat{\ladder}^{}_{Q} (r) &= \sum_{n_Q=1}^\infty \ \sum_{n_P,\nu_{\obar} , \nu_0 , \nu_1} \ket{n_P,n_Q-1\,;\,\nu_{\obar},\nu_0,\nu_1} \bra{n_P,n_Q\,;\,\nu_{\obar},\nu_0,\nu_1}_r ,
\end{align}
\end{subequations}
with the corresponding raising operators $\hat{\ladder}_P^\dagger(r)$, $\hat{\ladder}_Q^\dagger(r)$ being their conjugates.
When it comes to the fermions, some care is needed in order to ensure the proper realization of fermionic statistics, i.e., an ordering of creation operators acting on the global fermionic vacuum has to be prescribed for any populated modes.
A simple choice would be to order sites $1$ to $N$ from left to right, such that quarks at site $N$ are created first, then site $N-1$, and so on down to site 1, and within each lattice site $r$ the creation operators are ordered as $(\chi^\dagger_{\obar})^{\nu_{\obar}}(\chi^\dagger_{0})^{\nu_{0}}(\chi^\dagger_{1})^{\nu_{1}}$, like in Eq.~\eqref{eq: basis in term of creation operators}.
The appropriate definitions of the fermionic annihilation operators are then
\begin{subequations}
\begin{align}
    \hat{\chi}_{\obar}(r) &= \left( \prod_{r'=1}^{r-1} (-1)^{\hat{\nu}_{\obar}(r')+\hat{\nu}_0(r')+\hat{\nu}_1(r')} \right) \sum_{n_P , n_Q , \nu_0 , \nu_1 } \ket{n_P,n_Q\,;\,0 ,\nu_0 ,\nu_1} \bra{n_P,n_Q\,;\,1 ,\nu_0 ,\nu_1}_r, \label{eq: chi-1bar-global} \\
    \hat{\chi}_0 (r) &= \left( \prod_{r'=1}^{r-1} (-1)^{\hat{\nu}_{\obar}(r')+\hat{\nu}_0(r')+\hat{\nu}_1(r')} \right) \sum_{n_P , n_Q , \nu_{\obar} , \nu_1 } \ket{n_P,n_Q\,;\,\nu_{\obar} ,0 ,\nu_1} \bra{n_P,n_Q\,;\,\nu_{\obar} ,1 ,\nu_1}_r (-1)^{\nu_{\obar}}, \label{eq: chi-o-global} \\
    \hat{\chi}_{1}(r) &= \left( \prod_{r'=1}^{r-1} (-1)^{\hat{\nu}_{\obar}(r')+\hat{\nu}_0(r')+\hat{\nu}_1(r')} \right) \sum_{n_P , n_Q , \nu_{\obar} , \nu_0 } \ket{n_P,n_Q\,;\,\nu_{\obar} ,\nu_0 ,0} \bra{n_P,n_Q\,;\,\nu_{\obar} ,\nu_0 ,1}_r (-1)^{\nu_{\obar} + \nu_0}, \label{eq: chi_1_global} 
\end{align}
\end{subequations}
where $(-1)^{\hat{\nu}_f}$ is a conditional phase that can be equivalently expressed as $(-1)^{\hat{\nu}_f} = 1 - 2 \hat{\nu}_f$.
The creation operators $\hat{\chi}_{\obar}^\dagger(r)$,  $\hat{\chi}_0^\dagger(r)$, and  $\hat{\chi}_{1}^\dagger(r)$ are obtained by Hermitian conjugation.
The above definitions, when used in Eqs.~\eqref{eq: HM in LSH operators}-\eqref{eq: HI in LSH operators}, along with the diagonal functions appearing in $H_I$ (such as $\sqrt{1+\hnuo/(\hat{n}_P+1)}$) being defined as explained in the previous sections, give a complete description of the Hamiltonian matrix with respect to the LSH occupation-number basis.

Beyond the essential Abelian Gauss's law constraints, there are also superselection sectors corresponding to the different possible global quantum numbers, the discussion of which was given entirely in Sec.~\ref{subsec: super-selection}.


\section{Conclusion
\label{sec: discussion}
}
\noindent
Prior to this work, the LSH framework has been shown to have several advantages relative to other frameworks present in the literature for Hamiltonian simulation \cite{Davoudi:2020yln}.
One of the essential benefits offered by the LSH formalism is its concise solution to the problem of building the gauge invariant Hilbert space and keeping the dynamics inherently confined within that space \cite{Mathew:2022nep}.
Especially notable is the fact that, despite the 
considerably more involved intermediate steps necessary to derive the SU(3) formulation as compared with SU(2),
the end result still turns out to be a Hamiltonian and Hilbert space that is remarkably similar to that which is obtained in the simpler SU(2) theory:
(i) instead of one bosonic variable and two fermionic variables per site, one has two bosonic variables and three fermionic variables per site, and (ii) the gauge-matter interaction in Eq.~\eqref{eq: HI} naturally separates into three decoupled fermionic hopping terms, instead of two as in the case of SU(2).
Apart from these minimal extensions, the SU(3) LSH framework otherwise contains each and every important feature of the SU(2) framework that make it useful for practical applications.

The primary ingredients of the SU(3) LSH framework presented in this work can be summarized as:
(i) strictly SU(3)-invariant, site-local degrees of freedom consisting of two bosonic fluxes and three quark modes,
(ii) the Hamiltonian of (1+1)-dimensional Yang-Mills theory coupled to staggered quarks, written explicitly in terms of the LSH degrees of freedom and furthermore expressed with respect to an LSH occupation-number basis,
(iii) two Abelian constraints per link of the lattice,
and (iv) the super-selection rules for LSH dynamics.
We have also supplemented this framework with a pictorial representation of the states and operators in terms of directional gauge fluxes that may be sourced or sinked by the quarks.
Finally, the SU(3) LSH framework is numerically validated by a benchmark comparison with another formulation of the same theory that is currently being considered in the context of quantum simulation for QCD.

The structural similarity of the SU(3) framework to its SU(2) counterpart opens up the possibility of transferring many of the techniques developed for SU(2) over to the gauge group of QCD.
\begin{itemize}
    \item For example, past work that explained how to digitize physical states of SU(2) lattice gauge theory \cite{Raychowdhury:2018osk} is immediately applicable to the SU(3) framework now developed.
    \item In a recent work concerning digital quantum algorithms for SU(2) \cite{Davoudi:2022xmb}, it has been discovered that the Trotterization of a hopping term can be reduced from $2^3$ constituent steps (in the Kogut-Susskind formulation) to only two;
    the division of the SU(3) LSH hopping terms into three fermionic couplings suggests the possibility of an analogous reduction from $3^3$ costly subterms to only three.
    \item In the context of analog quantum simulation for the $1+1$-dimensional LSH Hamiltonian, it is indeed possible to simulate exact dynamics of the theory using a purely fermionic analog simulation platform, with nearest neighbour hopping terms, provided the simulating Hamiltonian can be set up to mimic the dynamics of fermionic LSH excitations.
    Such a proposal has already been made for the SU(2) theory in Ref. \cite{Dasgupta:2020itb} and may be extendable to simulating SU(3) LSH dynamics. 
    \item 
    An important feature of the SU(2) LSH framework was its strictly Abelian symmetries pertaining to conservation of a single type of flux along links;
    the same feature recurs for SU(3) with the only difference being that there are two types (directions) of Abelian flux now. Whether the calculations are quantum simulations or tensor network calculations, the continuity of these fluxes must be protected in the dynamics.
    A recently proposed symmetry protection scheme for the SU(2) LSH formulation \cite{Mathew:2022nep} should be generalizable to the SU(3) framework, which would be useful in simulating LSH dynamics using classical (such as tensor network) or quantum (analog or digital) architectures. 

    \item The purely fermionic formulation that was used as a benchmark in Sec.~\ref{sec: results} and described in Appendix \ref{app: purely fermionic formulation}, despite being extremely useful in 1+1 dimensions, has a nonlocal Hamiltonian, is difficult to apply to periodic boundary conditions, and simply has no higher-dimensional counterpart given the actual presence of transverse waves. The LSH formulation, by contrast, can accommodate both open and periodic boundary conditions, is local, and is expected to be generalizeable to higher dimensions like its SU(2) counterpart.
\end{itemize}

The driving motivation for the present work was the need for a convenient Hamiltonian framework for performing quantum simulations of lattice QCD.
With the framework introduced in this work, there are immediate research opportunities to explore benefits that should be offered by the SU(3) LSH formulation. Any such advantages should have immediate relevance toward the actual goal of quantumly simulating lattice QCD. 
In the near future, the development of quantum simulation algorithms or even tensor network algorithms for SU(3) gauge theory using the (1+1)-dimensional LSH framework would be interesting applications that can be pursued using just the present work as a basis.
This current work marks a concrete step forward towards the goal of making quantum simulations of QCD a reality, with some of the most important generalizations to be explored in coming years being those of higher spatial dimensions and the extension to multiple fermion flavors.

\section*{Acknowledgments}
\noindent
We acknowledge insightful discussions with Aniruddha Bapat, Zohreh Davoudi, and Emil Mathew.
SVK and JRS were supported by the U.S. Department of Energy’s Office of Science Early Career Award, under award DE-SC0020271 (PI Z. Davoudi). SVK further acknowledges support by the Maryland Center for Fundamental Physics at the University of Maryland, College Park. During the planning stages of the collaboration, IR was supported by the U.S.
Department of Energy’s Office of Science, Office of Advanced Scientific Computing Research, Quantum Computing Application Teams program, under fieldwork proposal number ERKJ347. Research of IR is supported by the Research Initiation Grant (BPGC/RIG/2021-
22/10-2021/01), OPERA award (FR/SCM/11-Dec-2020/PHY) from BITS-Pilani, and the Start-up Research Grant (SRG/2022/000972) from SERB, India. 

\bibliography{main.bib}
\appendix

\section{Purely fermionic formulation of SU(3) gauge theory
\label{app: purely fermionic formulation}
}
%
\begin{table}[t]
    \renewcommand{\arraystretch}{1.5}
    \centering
    \begin{tabular}{C{1cm}  C{2.5cm}  C{2cm} C{2.5cm}}
        \hline
        \hline
        \rule{0pt}{2ex}
        $\mathcal{F}$ & $(\mathcal{P}_f, \mathcal{Q}_f)$ & $d(\mathcal{P}_f, \mathcal{Q}_f)$ & Eigenvalue\\
        \hline
        0 & (0, 0) & 1 & 0.000 \\
        \hline
        \multirow{2}{*}{1} & \multirow{2}{*}{(1, 0)} & \multirow{2}{*}{3} & -0.387 \\
        & & & 1.721\\
        \hline
        \multirow{3}{*}{2} & \multirow{3}{*}{(0, 1)} & \multirow{3}{*}{3} & -1.535 \\
        & & & 0.868 \\
        & & & 3.333 \\
        \hline
        2 & (2, 0) & 6 & 1.333 \\
        \hline
        \multirow{4}{*}{3} & \multirow{4}{*}{(0, 0)} & \multirow{4}{*}{1} & -3.858 \\
        & & & -0.497 \\
        & & & 2.137 \\
        & & & 4.884 \\
        \hline
        \multirow{2}{*}{3} & \multirow{2}{*}{(1, 1)} & \multirow{2}{*}{8} & -0.081 \\
        & & & 2.747\\
        \hline
        \multirow{3}{*}{4} & \multirow{3}{*}{(1, 0)} & \multirow{3}{*}{3} & -2.562 \\
        & & & 1.089 \\
        & & & 4.140 \\
        \hline
        4 & (0, 2) & 6 & 1.333 \\
        \hline
        \multirow{2}{*}{5} & \multirow{2}{*}{(0, 1)} & \multirow{2}{*}{3} & -1.277 \\
        & & & 2.610\\
        \hline
        6 & (0, 0) & 1 & 0.000 \\
        \hline
    \end{tabular}
    \caption{Energy eigenvalues of the Hamiltonian in Eq.~\eqref{eq: ham-terms} with $N=2$ sites and $\mu= x = 1$ are shown in this table. Without any loss of generality, we  consider zero background flux, i.e. choosing $E^{\rm a}(R,0)=0$, $\forall a$ in Eq.~\eqref{eq: HE-ferm}. The eigenvalues are categorized according to according to total fermion number $\mathcal{F}$ and global charges $(\mathcal{P}_f,\mathcal{Q}_f)$ of the corresponding eigenstates. The dimension $d(\mathcal{P}_f,\mathcal{Q}_f)$ of the irrep $(\mathcal{P}_f,\mathcal{Q}_f)$ is given by Eq.~\eqref{eq: dim of irrep}, which is also the degeneracy of the eigenvalues in the purely fermionic formulation. Thus, each eigenvalue when counted with its corresponding degeneracy factor add up to $8^N=64$, which is the dimension of the Hilbert space of this system in the purely fermionic formulation. In the LSH framework, each eigenvalue in each global symmetry sector  has only one eigenstate as explained in Sec.~\ref{subsec: result-numerics}. \label{tab: Eigenvalues}}
\end{table}
The LSH Hamiltonian along with the associated Hilbert space presented in section \ref{sec: LSH-framework} reformulates the Kogut-Susskind Hamiltonian in Eq.~\eqref{eq: KS-ham} in terms of Schwinger bosons and staggered fermions. However, there are other formulations of the same Hamiltonian in terms of different degrees of freedom~\cite{Farrell:2022wyt, Farrell:2022vyh, Atas:2022dqm}. We present one such formulation in 1+1 dimensions with open boundary condition in which the gauge degrees of freedom are removed by working with a particular gauge choice, and the resulting Hamiltonian has only colored fermionic degrees of freedom. This is known as the gauge fixed formulation or purely fermionic formulation.

Under the Gauss's law constraint in Eq.~\eqref{eq: gauss-law}, the physical Hilbert space of the SU(3) Kogut-Susskind Hamiltonian in $1+1$ dimensions has no dynamical gauge degrees of freedom except for the possible closed flux loops in the case of a periodic boundary condition. Thus, imposing the open boundary condition, where the closed loops are absent, and fixing the incoming right chromo-electric flux at the first site, denoted by $E^{\rm a}(R,0)$, the chromo-electric field in the entire lattice is completely determined for a given fermionic configuration. Equation~\eqref{eq: gauss-law} can then be used with a suitable choice of gauge fixing to eliminate all the intermediate gauge-link degrees of freedom as shown below. As a result, the SU(3) Kogut-Susskind Hamiltonian acting on the physical Hilbert space can be expressed in terms of purely fermionic degrees of freedom and the left boundary ${E}^{\rm a}(R,0)$ flux. The derivation shown here is extension of the similar formulation for SU(2) theories given in Refs.~\cite{Atas:2021ext,Davoudi:2020yln}. 

Consider the Hamiltonian in Eq.~\eqref{eq: KS-ham} with the following gauge transformation of the fermionic fields:
\begin{equation}
    \psi(r)\to\psi'(r) = \left[\prod_{y<r}U(y)\right] \psi(r),~~\quad~~
    \psi^\dagger(r)\to\psi^{\dagger'}(r)= \psi^\dagger(r)\left[\prod_{y<r}U(y)\right]^\dagger,
    \label{eq: gauge-transform-ferm}
\end{equation}
where $\prod_{y<r}U(y) = U(1)U(2)\cdots U(r-1)$. The corresponding gauge transformation of the gauge links is given by
\begin{equation}
    U(r)\to U'(r) =  \left[\prod_{y<r}U(y)\right]U(r)\left[\prod_{z<r+1}U(z)\right]^\dagger,
    \label{eq: gauge-transform-U}
\end{equation}
such that the gauge-matter interaction term, $H_{I}$, in Eq.~\eqref{eq: KS-ham} remains invariant. Since the gauge links satisfy the unitarity condition $U(r)U^\dagger(r)=\mathds{1}_{3\times3}$, $U'(r)$ in Eq. \ref{eq: gauge-transform-U} simplifies to $U'(r)=\mathds{1}_{3\times3}$. This transforms $H_{I}$ to
\begin{equation}
    H^{\text(F)}_{I} = \sum_{r=1}^{N-1}\psi^{\dagger'}(r)\cdot \psi'(r+1) + \text{H.c.},
    \label{eq: HI-ferm}
\end{equation}
where the superscript $F$ denotes the purely fermionic formulation. Similarly, the mass term, $H_M$, in Eq.~\eqref{eq: KS-ham} is given by
\begin{equation}
    H^{\text(F)}_M = \mu \sum_{r=1}^{N}(-1)^r\psi^{\dagger'}(r)\cdot \psi'(r).
    \label{eq: HM-ferm}
\end{equation}
Finally, the electric field energy term, $H_E$, in Eq.~\eqref{eq: KS-ham} can be re-expressed in terms of the fermionic field and the left boundary electric flux, $E^{\rm a}(R,0)$, using the Gauss's law stated in Eq.~\eqref{eq: gauss-law}. For a physical state that satisfies Eq.~\eqref{eq: gauss-law}, the electric field operator to the right of any site $r>1$ can be evaluated by iteratively using the Gauss's law operator defined in Eq.~\eqref{eq: gauss-op}:
\begin{equation}
    \hat{E}^{\rm a}(L,r) = E^{\rm a}(R,0) + \sum_{r'=1}^{r-1} \psi^{\dagger}(r') T^{\rm a} \psi(r').
    \label{eq: Ex-gauge-fixed}
\end{equation}
Thus, $H_E$ in the purely fermionic formulation is given by
\begin{equation}
    H^{\text(F)}_E = \sum_{r=2}^{N-1}\sum_{{\rm a}=1}^8 \left[E^{\rm a}(R,0) + \sum_{r'=1}^{r-1} \psi^{\dagger}(r') T^{\rm a} \psi(r')\right]^2.
     \label{eq: HE-ferm}
\end{equation}
As a result, the final Hamiltonian given by Eqs.\eqref{eq: HI-ferm},~\eqref{eq: HM-ferm}, and~\eqref{eq: HE-ferm} has only the fermionic degrees of freedom and $E^{\rm a}(R,0)$ acts as the background chromo-electric field. This Hamiltonian is identical to the original Kogut-Susskind Hamiltonian only in the physical Hilbert space. Thus, any state in the physical Hilbert space in this formulation, $|\Psi\rangle^{(F)}$, can be characterized by its fermionic occupation number at each site:
\begin{equation}
    |\Psi\rangle^{(F)}=\goldieotimes_{r=1}^{N}|f_1,f_2,f_3\rangle_{r}= \prod_{r=1}^{N}(\psi^\dagger_1(r))^{f_1(r)}(\psi^\dagger_2(r))^{f_2(r)}(\psi^\dagger_3(r))^{f_3(r)}\goldieotimes_{r'=1}^{N}|0,0,0\rangle_{r'}
    \label{eq: Fermionic basis state}
\end{equation}
where the subscript $r$ on the state vector denotes the local state at site $r$, $f_{i}(r)=0$ or $1$ refers to the occupation number of the (anti)matter field at site $r$ with color index $i=1,2,3$, and the vacuum state $|0,0,0\rangle_r$ is defined by $\psi^\alpha |0,0,0\rangle_r =0 $ for $\alpha = 1,2,3$.

Equation~\eqref{eq: Fermionic basis state} implies that the dimension of the fermionic Hilbert space of an $N$ site lattice in this formulation is given by $8^N$. 
Furthermore, the fermionic creation operators, $\psi^\dagger_\alpha(r)$, transform in the $(1,0)$ irrep under the local SU(3) gauge group as indicated in Table~\ref{tab: prepotential irreps}, and satisfy anti-commutation relations in Eq.~\eqref{eq: ferm_anticomm}.
Thus, the states $|\Psi\rangle^{(F)}$ with total fermion number $\mathcal{F}$ has $\mathcal{F}$ color indices, and their linear combinations with appropriate Clebsh-Gordon coefficients form the multiplets that transform according to different $(\mathcal{P}_f,\mathcal{Q}_f)$ irreps under the global SU(3) transformations as discussed in Sec.~\ref{subsec: result-numerics}.
As an example, in Table~\ref{tab: Eigenvalues}, we show the degeneracy factors for energy eigenvalues for vanishing left boundary flux as shown in Fig.~\ref{subfig: eigenavlues}. 
The eigenvalues are categorized according to $\mathcal{F}$ and $(\mathcal{P}_f,\mathcal{Q}_f)$ global charges of their eigenstates with degeneracy factor for each eigenvalue given by $d(\mathcal{P}_f,\mathcal{Q}_f)$, which is $d(\Delta\mathcal{P},\Delta\mathcal{Q})$ for $(\mathcal{P}_0,\mathcal{Q}_0)=(0,0)$.

\section{Additional LSH operator factorizations
\label{app: extra factorizations}}

 In addition to the local SU(3) singlet operator factorizations that are directly relevant to expressing the Hamiltonian (see Eqs.~\eqref{eq: on site operator factorinzation}), one may wish to have a complete suite of on-site SU(3) singlet factorizations for constructing the local Hilbert space as originally given in Eqs.~\eqref{eq: LSH states unnormalized def}-\eqref{eq: Unnormalized LSH-basis-state}.
 The following factorizations supplement Eqs.~\eqref{eq: on site operator factorinzation} to complete that set:
\begin{align}
    \ApBp &\quad\mapsto\quad \hat{\ladder}_P^\dagger \sqrt{\hat{n}_P+1} \sqrt{\hat{n}_P+3-\delta_{\hnum, \hat{\nu}_1}} \sqrt{ 1 + \frac{1}{\hat{n}_P+\hat{n}_Q+3-\delta_{\hnui, \hnum}\delta_{\hnum , \hat{\nu}_1}} } \\
    \BpAp &\quad\mapsto\quad \hat{\ladder}_Q^\dagger \sqrt{\hat{n}_Q+1} \sqrt{\hat{n}_Q+3-\delta_{\hnui, \hnum}} \sqrt{ 1 + \frac{1}{\hat{n}_P+\hat{n}_Q+3-\delta_{\hnui, \hnum}\delta_{\hnum , \hat{\nu}_1}} } \\
    \tfrac{1}{2} \psi^\dagger \cdot \psi^\dagger \wedge A(\obar)^\dagger &\quad\mapsto\quad \hat{\chi}_{\obar}^\dagger \hat{\chi}_0^\dagger ( \hat{\Lambda}_P^+ )^{\hat{\nu}_1} \sqrt{\hat{n}_P+2-\hat{\nu}_1} \sqrt{1+\frac{1-\hat{\nu}_1}{\hat{n}_P+\hat{n}_Q+2}} \\
    \tfrac{1}{2} \psi^\dagger \cdot \psi^\dagger \wedge A(1)^\dagger &\quad\mapsto\quad \hat{\chi}_0^\dagger \hat{\chi}_1^\dagger ( \hat{\Lambda}_Q^+ )^{\hat{\nu}_{\obar}} \sqrt{\hat{n}_Q+2-\hat{\nu}_{\obar}} \sqrt{1+\frac{1-\hat{\nu}_{\obar}}{\hat{n}_P+\hat{n}_Q+2}} \\
    \psi^\dagger \cdot A(\obar)^\dagger \wedge A(1)^\dagger &\quad\mapsto\quad \hat{\chi}_0^\dagger (\hat{\Lambda}_P^+)^{\hat{\nu}_1} (\hat{\Lambda}_Q^+)^{\hat{\nu}_{\obar}} \sqrt{\hat{n}_P+2-\hat{\nu}_1} \sqrt{\hat{n}_Q+2-\hat{\nu}_{\obar}} \nonumber \\
    & \qquad \qquad \times \sqrt{1+\frac{1}{\hat{n}_P+\hat{n}_Q+2+\hat{\nu}_{\obar}+\hat{\nu}_1-\hat{\nu}_{\obar} \hat{\nu}_1}} 
\end{align}
%
\end{document}